\documentclass[9pt,twoside,twocolumn,final]{pnas-new}
 
\usepackage[utf8]{inputenc}
 
\def\longformpaper{}

\usepackage{bbm}
\usepackage{amsmath,amssymb}
\usepackage{subfigure}

\usepackage{hyperref}
\hypersetup{
 colorlinks=true,
 citecolor=blue,
 linkcolor=blue,
 urlcolor=blue
}	

\usepackage[normalem]{ulem}

\templatetype{pnasresearcharticle}

\usepackage{float}
\usepackage{tikz}
\usetikzlibrary{shapes.geometric,arrows.meta,arrows}

\title{How Market Ecology Explains Market Malfunction}

\author[a,b,1]{Maarten P. Scholl}
\author[b,1]{Anisoara Calinescu}
\author[a,c,d,2]{J. Doyne Farmer} 
\affil[a]{Institute for New Economic Thinking, Oxford Martin School, University of Oxford}
\affil[b]{Computer Science Department, University of Oxford}
\affil[c]{Mathematical Institute, University of Oxford}
\affil[d]{Santa Fe Institute}

\leadauthor{Scholl} 

\authordeclaration{The authors declare no conflict of interest.}
\correspondingauthor{\textsuperscript{1} E-mail: maarten.scholl@cs.ox.ac.uk}


\begin{abstract}
Standard approaches to the theory of financial markets are based on equilibrium and efficiency.  Here we develop an alternative based on concepts and methods developed by biologists, in which the wealth invested in a financial strategy is like the abundance of a species. We study a toy model of a market consisting of value investors, trend followers and noise traders.  We show that the average returns of strategies are strongly density dependent, i.e. they depend on the wealth invested in each strategy at any given time. In the absence of noise the market would slowly evolve toward an efficient equilibrium, but the statistical uncertainty in profitability (which is adjusted to match real markets) makes this noisy and uncertain.  Even in the long term, the market spends extended periods of time away from perfect efficiency. We show how core concepts from ecology, such as the community matrix and food webs, give insight into market behavior.  The wealth dynamics of the market ecology explain how market inefficiencies spontaneously occur and gives insight into the origins of excess price volatility and deviations of prices from fundamental values.
\end{abstract}
\significancestatement{
We develop the mathematical analogy between financial trading strategies and biological species and show how to apply standard concepts from ecology to financial markets.  We analyze the interactions of three stereotypical trading strategies in ecological terms, showing that they can be competitive, predator-prey or mutualistic, depending on the wealth invested in each strategy.  The  deterministic dynamics suggest that the system should evolve toward an efficient state where all three strategies make the same average returns. However, this happens so slowly, and the evolution is so noisy, that there are large fluctuations away from the efficient state, causing bursts of volatility and extended periods where prices deviate from fundamental values. This provides a conceptual framework that gives insight into the reasons why markets malfunction.   
}

\doi{\url{www.pnas.org/cgi/doi/10.1073/pnas.XXXXXXXXXX}}

\begin{document}

\maketitle
\thispagestyle{firststyle}
\ifthenelse{\boolean{shortarticle}}{\ifthenelse{\boolean{singlecolumn}}{\abscontentformatted}{\abscontent}}{}
    \vspace{-1em}
    
\dropcap{W}hy do markets malfunction?  According to the theory of market efficiency, markets always function perfectly.  Prices always reflect fundamental values and they only change when there is new information that affects fundamental values.  Thus, by definition, any problems with price setting are caused by factors outside the market. Empirical evidence suggests otherwise.  Large price movements occur even when there is very little new information \cite{Cutler1989} and prices often deviate substantially from fundamental values \cite{Shiller1989}.  This indicates that, to understand how and why markets malfunction, we need to go beyond the theory of market efficiency.  
    
Here we build on earlier work \cite{Farmer2002MarketEvolution,Lebaron2002BuildingMarketc, lo2004,Hens2009,Farmer2013,Beale2019DynamicsMarkets} and develop the theory of market ecology, which provides just such an alternative.  This approach borrows concepts and methods from biology and applies them to financial markets.  Financial trading strategies are analogous to biological species. Plants and animals are specialists that evolve to fill niches that provide food; similarly, financial trading strategies are specialized decision making rules that evolve to exploit market inefficiencies.  Trading strategies can be classified into distinct categories, such as technical trading, value investing, market making, statistical arbitrage and many others. The capital invested in a strategy is like the population of a species. Trading strategies interact with one another via price setting and the market evolves as the wealth invested in each strategy changes through time, and as old strategies fail and new strategies appear. 
    
The theory of market ecology emerges from the inherent contradictions in the theory of market efficiency. A standard argument used to justify market efficiency is that competition for profits by arbitrageurs should cause markets to rapidly evolve to an equilibrium where it is not possible to make excess profits based on publicly available information. But if there are no profits to be made, there are no incentives for arbitrageurs, so there is no mechanism to make markets efficient. This paradox suggests that, while markets may be efficient in some approximate sense, they cannot be perfectly efficient \cite{Grossman1980OnMarkets}.  In contrast, under the theory of market ecology, trading strategies exploit market inefficiencies but, as new strategies appear and as the wealth invested in each strategy changes, the inefficiencies change as well.  To understand how the market functions, it is necessary to understand how each strategy affects the market and how the interactions between strategies cause market inefficiencies to change with time.  The theory of market ecology naturally addresses a different set of problems than the theory of market efficiency, and can be viewed as a complement rather than a substitute.

Our study here builds on a large body of work on agent-based models of financial markets, e.g. \cite{Arthur1997AssetMarket, brock1997, Lux1999ScalingMarket, Chiarella1992TheBehaviour}. The theory of market ecology provides a conceptual framework for understanding such models. Our goal is not to construct a better model of financial markets, but rather to show how ideas from ecology can be used to interpret market phenomena and predict market behavior.
    
Here we study a stylized toy market model with three trading strategies.  We approach the problem in the same way that an ecologist would study three interacting species.  We study how the average returns of the strategies depend on the wealth invested in each strategy, how their wealth evolves through time under reinvestment, and how their endogenous time evolution causes the market to malfunction.


We show that, with realistic parameters, evolution toward market efficiency is very slow.  The expected deviations from efficiency are in some sense small, but they persist even in the long term, and cause extended deviations from fundamental values and excess volatility (which in extreme cases becomes market instability).  Our study provides a simple example of how analyzing markets in these terms and tracking market ecology through time could give regulators and practitioners better insight into market behavior. 

    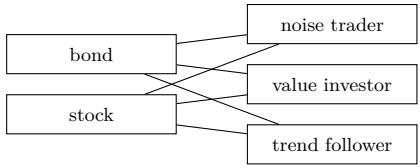
\begin{figure}
        \vspace{-1em}
    	\centering
    	\caption[Diagram 1.]{The three trading strategies correspond to noise traders, value investors and trend followers.  They invest their capital in a stock and a bond.  The mixture for each strategy changes with time as strategies accumulate wealth based on their historical performance.
    	}
    	\label{diagram:marketstructure}
    	\tikzstyle{block}  = [rectangle, draw, text width=8em   , text centered, minimum height=2em] 
    	\tikzstyle{invis}  = [fill=none,draw=none, text centered, minimum height=2em]
    	\tikzstyle{line}   = [draw, -latex']
    	\begin{tikzpicture}[node distance=4cm,auto,scale=0.8, every node/.style={transform shape}]
    	\node at(0, -0.5)   [block] (asset0)     {bond};
    	 
    	\node at(0, -1.5)  [block] (asset1)     {stock};
    	\node at(4.0, 0)   [block] (agent0)     {noise trader};
    	\node at(4.0,-1)   [block] (agent1)     {value investor};
    	\node at(4.0,-2)   [block] (agent2)     {trend follower}; 
	    
        	\path [line, -] (asset0) -- (agent0);
        	\path [line, -] (asset0) -- (agent1);
        	\path [line, -] (asset0) -- (agent2);
        	
        	\path [line, -] (asset1) -- (agent0);
        	 \path [line, -] (asset1) -- (agent1);
        	 \path [line, -] (asset1) -- (agent2);
    	\end{tikzpicture}\vspace{-1em}
    \end{figure}
   
   \subsection{Model Description}

The structure of the model is schematically summarized in Figure \ref{diagram:marketstructure}.  There are two assets, a stock and a bond. The bond trades at a fixed price and yields $r=1\%$ annually in the form of coupon payments that are paid out continuously. The stock pays a dividend $\boldsymbol{D}(t)$ at each time step that is modeled as a discrete-time autocorrelated geometric Brownian motion, of the form    
    
    \begin{equation}
    \begin{split}\label{label:multivariate_geometric_brownian_motion_autocorrelation}
            \boldsymbol{D}(t) &= D(t) + g \boldsymbol{D}(t-1) + \sigma \boldsymbol{D}(t-1) \boldsymbol{U}({t}), \\ 
            \boldsymbol{U}(t) &=  \omega \boldsymbol{U}(t-2) + (1-\omega^2)\boldsymbol{Z}(t). \\
    \end{split}
    \end{equation}
where $g$ is the average rate of dividend payments, $\sigma$ is the variance, $\omega$ is the autocorrelation parameter of the process, and $Z$ is a  standard Wiener process. We choose parameters so that one time step is roughly equal to a day. 
We use estimates from market data by LeBaron \cite{Lebaron2001EmpiricalMarket}, taking $g=2\%$ per year for the growth rate of the dividend with a volatility of $\sigma=6\%$. (See reference \cite{Fama1988DividendReturns}, for example, for a review of the empirical evidence on dividends).
   
 We use market clearing to set prices.  The stock has a fixed supply $Q$, but the excess demand $E(t)$ for the stock by each trading strategy varies in time.  We allow the trading strategies to take short positions and to use leverage (i.e. to borrow in order to take a position in the stock that is larger than their wealth).  We impose a strategy-specific leverage limit $\lambda^*$.  Because we use leverage and because the strategies can have demand functions with unusual properties, market clearing is not always straightforward -- see  Materials and Methods.

 The size of a trading strategy is given by its {\it wealth} $W(t)$, i.e. the capital invested in it at any given time.  In ecology this corresponds to the population of a species, which is also called its abundance.  Unless otherwise stated, the wealth of each strategy varies proportional to its cumulative performance. Letting $\pi_{i}(t)$ be the return of strategy $i$ at time $t$, the wealth changes according to 
 \begin{equation}
 \label{wealth_reallocation}
   W_i(t+1) = (1 + f\pi_{i}(t)) W_i(t).  
 \end{equation} 
 The {\it reinvestment rate} $f$ models investor flows of capital. The default value $f=1$ corresponds to passive reinvestment, and $f>1$ means that profitable strategies attract additional capital and unprofitable strategies lose additional capital.

 A trading strategy is defined by its trading signal $\phi(t)$, which can depend on the price $p(t)$ and other variables, such as dividends and past prices.  We modify $\phi$ by a $\tanh$ function to ensure that the excess demand is bounded and differentiable.  A strategy's excess demand for the stock is
  \begin{align}\label{eqn:excess_demand}
         \begin{split} 
          E(t) &=~\frac{W(t)\lambda^*}{p(t)} \left(\tanh{  \left(c \cdot \phi(t)\right)} + \frac{1}{2}\right) - S(t-1),
          \end{split}
 \end{align} 
 where $S(t-1)$ is the number of shares of the stock held at the previous time step. The parameter $c > 0$ determines the aggressiveness of the response to the signal $\phi$, and is strategy specific. When the signal of the strategy is zero, the agent is indifferent between the stock and the bond and splits its portfolio equally between the two (hence the term of $1/2$).
The leverage $\lambda(t)$ of a strategy at any given time is $\frac{p(t)S(t)}{K(t)} =  \lambda^* | \tanh{\left( c \cdot \phi(t)\right)} + \frac{1}{2} |$. This equality holds when the market clears.
 
\subsection{Investment Strategies}\label{section:strategies}

We study three typical trading strategies, which we call value investors, trend followers and noise traders. We intentionally make all strategies boundedly rational, i.e. they work from limited information and their strategies are not optimal. We use a representative agent hypothesis, treating each strategy as though it were only used by a single fund; however, these should be thought of as representing all investors using these strategies.  We now describe each strategy in turn.

 {\bf Value Investors}\label{section:fundamentalvalueinvestor} observe the dividend process and use a model to derive the value of the stock. They seek to hold more of the stock when it is undervalued and hold more of the bond when the stock is overvalued.  The parameters of their model are estimated based on historical dividends.  

 The fundamental value $V(t)$ of the stock at $t=0$ is given as the discounted expected future dividends:
    
    \begin{equation}\label{eq:gordongrowthmodel}
        V(0) = \mathbb{E}\left[ \sum_{t=1}^{\infty} \frac{D(t) }{(1+k)^t}\right] 
    \end{equation} 
The parameter $k$ is a discount rate called the \emph{required rate of return}, with $k \geq r$. $k$ is the sum of the risk-free rate $r$ and a risk-premium all the investors in our model expect for the additional risks associated with the stock. We  follow \cite{lebaron2006} and use a fixed discount rate $k=2\%$, based on the average rate of return implied by historical data. 
  
  The valuation $V^{\mathtt{VI}}(t)$ used by the value investors is made by estimating the mean growth rate $g$ from past data using the classical dividend discount model \citep{Gordon} $\mathbb{E}[D(t)] = D(0)(1+g)^t$. This ignores the autocorrelation coefficient $\omega$ and so in general $V^{\mathtt{VI}}(t) \ne V(t)$.
 
We define the trading signal for the value investor as the difference in log prices between the estimated fundamental value ${V^{\mathtt{VI}}}(t)$ and the market price.
    \begin{equation}
    \label{equation:fundamental_value_investor_signal}
        \phi_{\mathtt{VI}}(t) = \log_2 V^{\mathtt{VI}}(t) - \log_2 p(t)  
    \end{equation}
This strategy enters into a long position when the proposed price is lower than the estimated fundamental value and enters into a short position when the proposed price is higher than the estimated fundamental value. The use of the base two logarithm means that the value investor employs all of its assets when the stock is trading at half the perceived value \cite{Black1974TheReturns}.   
      
{\bf Trend Followers}\label{section:trendfollower}
expect that historical trends in returns continue into the short term future.  Several variants exist in the literature, including the archetypal trend follower that we use here \cite{brock1997, brock1998, joshi2002, Farmer2002MarketEvolution}.  There is evidence to suggest that trend-based investment strategies are profitable over long time horizons, and reference \cite{Asness2013} argues that investors earn a premium for the liquidity risk associated with stocks with high momentum (momentum trading is a synonym for trend following).

The trend following strategy extrapolates the trend in price between recently realized prices $p(t-2)$ and $p(t-1)$ time steps in the past. It always buys when prices trend upwards and sells when they trend downwards.
    \vspace{-0.5em}
    \begin{equation}
    \label{equation:trendfollowingsignal}
\phi_{\mathtt{TF}}(t) =  \log_2{p(t-2)} - \log_2{p(t-1)}.
    \end{equation}
 The trend followers' demand is a decreasing function of price.  Unlike the value investor, who cannot observe autocorrelations in dividends, the trend follower can exploit the autocorrelation that dividends impart to prices. 

{\bf Noise Traders}\label{section:noisetraders} represent non-professional investors who do not track the market closely.  Their transactions are mostly for liquidity, but they are also somewhat aware of value, so they are slightly more likely to buy when the market is undervalued and slightly more likely to sell when the market is overvalued (this is necessary so that their positions do not diverge over long periods of time).   
    
The signal function of our noise traders contains the product of the value estimate $V^{\mathtt{VI}}(t)$ (which is the same as for the value investors) and a stochastic component $X(t)$, \vspace{-1em}
    
    \begin{equation}\label{equation:noise_trader_univariate_sde}
    \begin{split}
    	 \phi_{\mathtt{NT}}(t) = \log_2 X(t)V^{\mathtt{VI}}(t) - \log_2 p(t).
    \end{split}
    \end{equation}
The noise process $X(t)$ is a discretized Ornstein-Uhlenbeck process, which has the form
   \begin{equation}\label{equation:ornstein_uhlenbeck_sde}
        X(t) = X(t) + \rho(\mu - X(t-1)) + \gamma \epsilon
    \end{equation}
This process reverts to the long term mean $\mu=1$  with reversion rate $\rho = 1- \sqrt[\leftroot{-2}\uproot{2}6\times 252]{0.5}
$, meaning the noise has a half life of 6 years, in accordance with the values estimated by Bouchaud \cite{Bouchaud2017}. $\epsilon$ is a standard normal random variable and $\gamma=12\%$ is a volatility parameter,   chosen so that noise traders generate volatility in excess of the volatility of dividends, matching the level observed in real world markets. 

The parameters of the model are summarized in Table~\ref{table:default_params}.  We have chosen them for an appropriate tradeoff between realism and conceptual interest, e.g. so that each strategy has a region in the wealth landscape where it is profitable.

    \section*{Results}
         
    \phantomsection
    \subsection{Density Dependence}\label{section:density_dep}
    \begin{figure}[htb!]
        \centering
        \includegraphics[width=\linewidth]{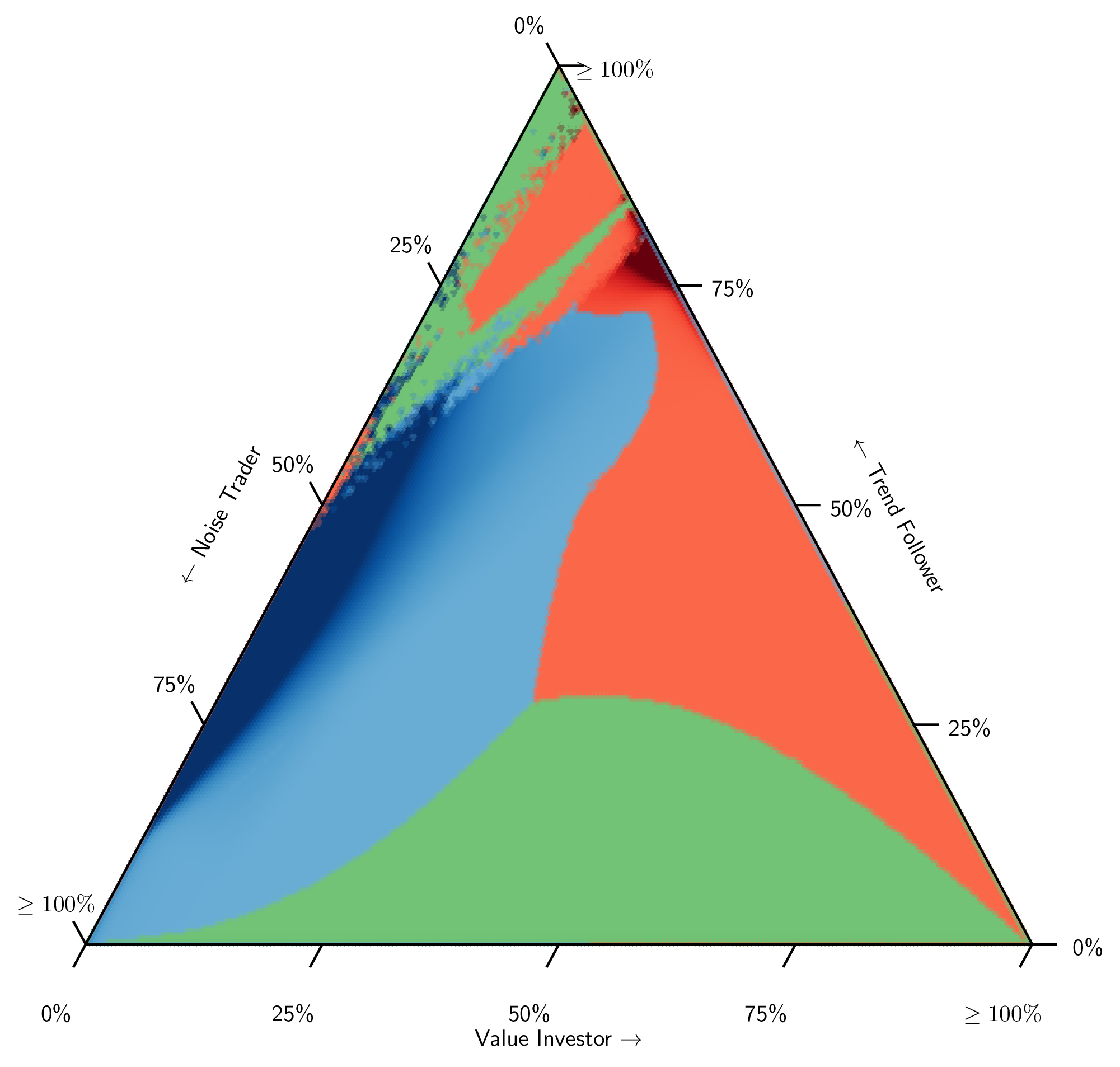}
        \includegraphics[width=\linewidth]{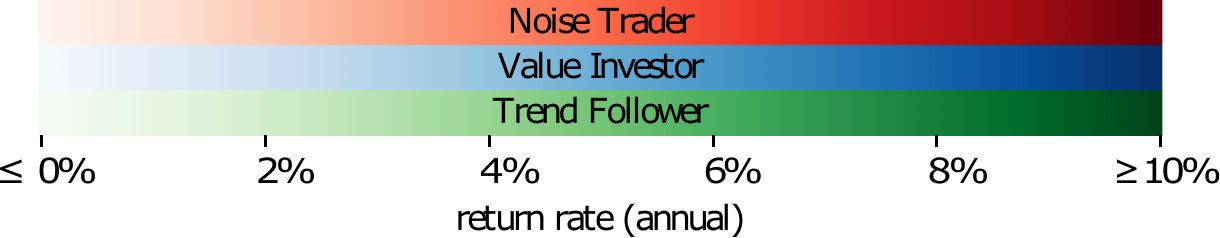}
        \includegraphics[width=\linewidth]{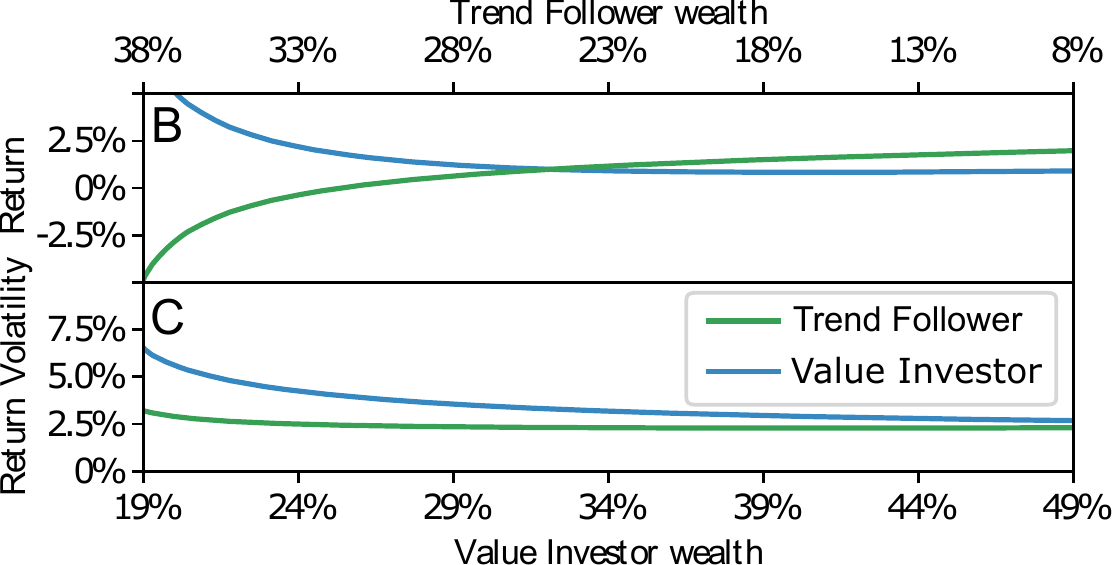}
         
        \caption{{\bf The profitability of the dominant strategy in the wealth landscape.} Panel {\bf A} is a ternary plot which displays the returns achieved by the strategy with the largest return. The axes correspond to the relative wealth invested in each strategy.  The top corner is pure trend followers, the left corner pure noise traders, and the right corner pure value investors. The color indicates the strategy with the highest returns at a given relative wealth vector $\boldsymbol{w}$. The regions colored in red correspond to the noise traders, blue regions to value investors, and green to the trend followers. The intensity of the color indicates the size of the average return.  The upper box of panel {\bf B} shows the average returns to value investors (blue) and trend followers (green) while holding the noise trader wealth at its equilibrium value of $42\%$.  The lower box shows the volatility in the returns of each strategy.  The horizontal axis is the relative wealth of the trend follower (top) and value investor (bottom).
        }\label{figure:performance_dd_sharpe}
    \end{figure}\vspace{-1em}

An ecosystem is said to have density dependence if its characteristics depend on the population sizes of the species, as is typically the case. Similarly, a market ecosystem is density dependent if its characteristics depend on the wealths of the strategies (both its own wealth and that of other strategies).  The toy market ecosystem that we study here is strongly density dependent. 

When the core ideas in this paper were originally introduced in reference \cite{Farmer2002MarketEvolution}, prices were formed using a market impact function, which translates the aggregate trade imbalance at any time into a shift in prices.  This can be viewed as a local linearization of market clearing.  The use of a market impact function suppresses density dependence and neglects non-linearities that are important for understanding market ecology.  

Using market clearing we see strong density dependence.  This is evident in Figure~\ref{figure:performance_dd_sharpe}, which shows which strategy makes the highest profits as a function of the relative size of each of the three strategies. To control the size of each strategy we turn off reinvestment, and instead replenish the wealth of each strategy at each step as needed, to hold it constant.  We then systematically vary the wealth vector $\boldsymbol{W}=(W_{\mathtt{NT}}, W_{\mathtt{VI}}, W_{\mathtt{TF}})$.  We somewhat arbitrarily let the total wealth $W_T = W_{\mathtt{NT}} + W_{\mathtt{VI}} +  W_{\mathtt{TF}} = 3 \times 10^8$.  For convenience of interpretation we plot the relative wealth $w_i(t) = W_i/W_T$. The results shown are averages over many long runs; to avoid transients we exclude the first 252 time steps, corresponding to one trading year.

Roughly speaking, the profitability of the dominant strategy divides the wealth landscape into four distinct regions.  Trend followers dominate at the bottom of the diagram, where their wealth is small.  Value investors dominate on the left side of the diagram, where their wealth is small, and noise traders dominate on the right side of the diagram, where their wealth is small.  There is an intersection point near the center where the returns of all three strategies are the same. In addition, there is a complicated region at the top of the diagram, where no single strategy dominates.  The turbulent behavior in this region comes about because the wealth invested by trend followers is large and the price dynamics are unstable. We do not regard this region as realistic, except perhaps in rare extreme market conditions.
    
A quantitative snapshot of the average returns and volatility is given in Figure~\ref{figure:performance_dd_sharpe}B, where we hold the size of the noise traders constant at $42\%$, corresponding to their wealth at the intersection point, and vary the wealth of the value investors and trend followers.   The average return to both trend followers and value investors decreases monotonically as their wealth increases.  The volatility of the returns of both strategies, in contrast, is a monotonic function of the wealth of the trend followers alone -- higher trend follower wealth implies higher volatility.   Although this is not shown here, the average return of the value investors increases strongly with the wealth of the noise traders; in contrast, the average return of the trend followers is insensitive to it.   

\subsection{Adaptation and the slow approach to market efficiency}\label{section:evolution}

\sidecaptionvpos{figure}[
    \begin{figure*}
        \centering
        \begin{subfigure}
            \centering \raisebox{3mm}{\put(0,0){\includegraphics[width=0.465\textwidth]{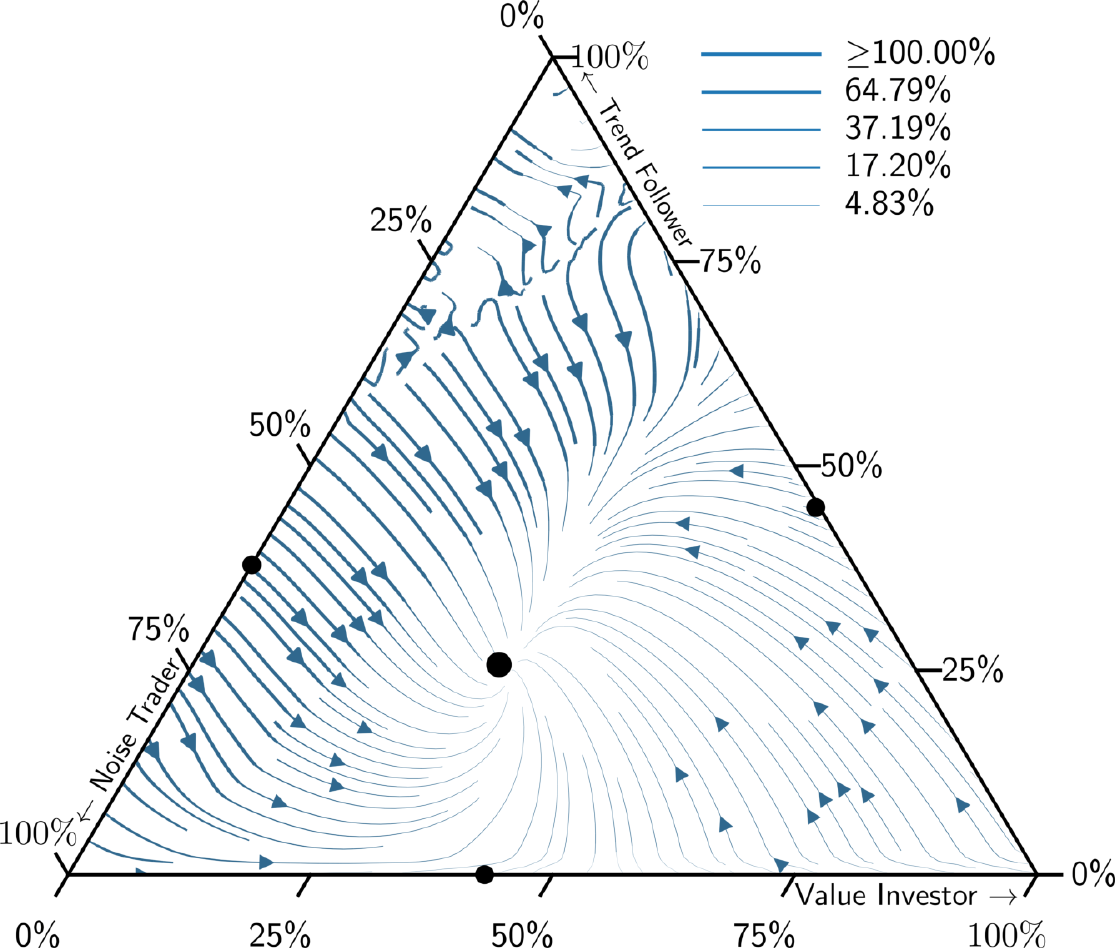}}\put(10,180){\Huge A}}
        \end{subfigure}
        \hfill
        \begin{subfigure}
            \centering{\includegraphics[width=0.45\textwidth]{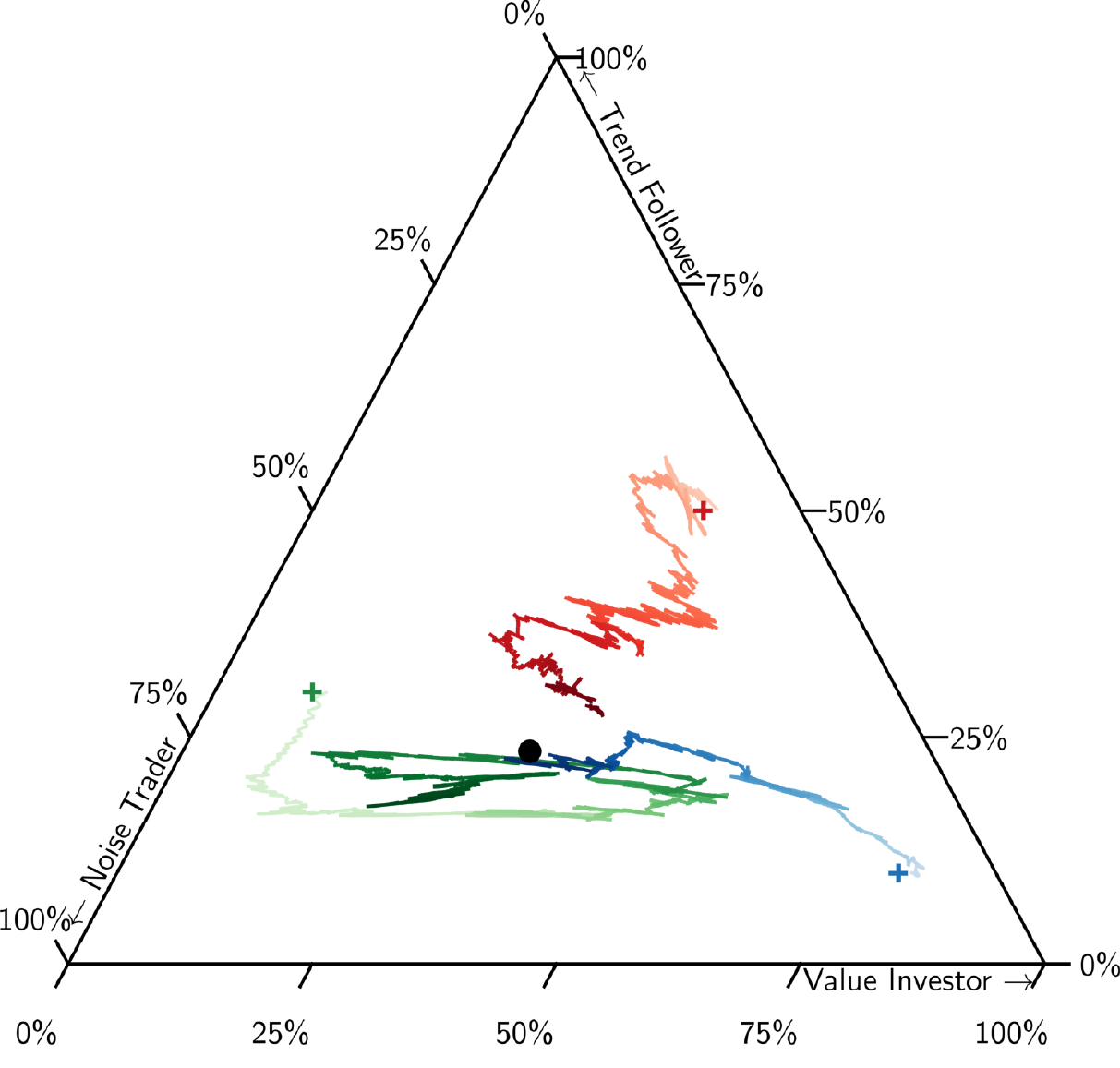}}\put(-10,180){\Huge B}
        \end{subfigure}
        \vskip\baselineskip
        \begin{subfigure}  
            \centering \put(0,0){\includegraphics[width=0.445\textwidth]{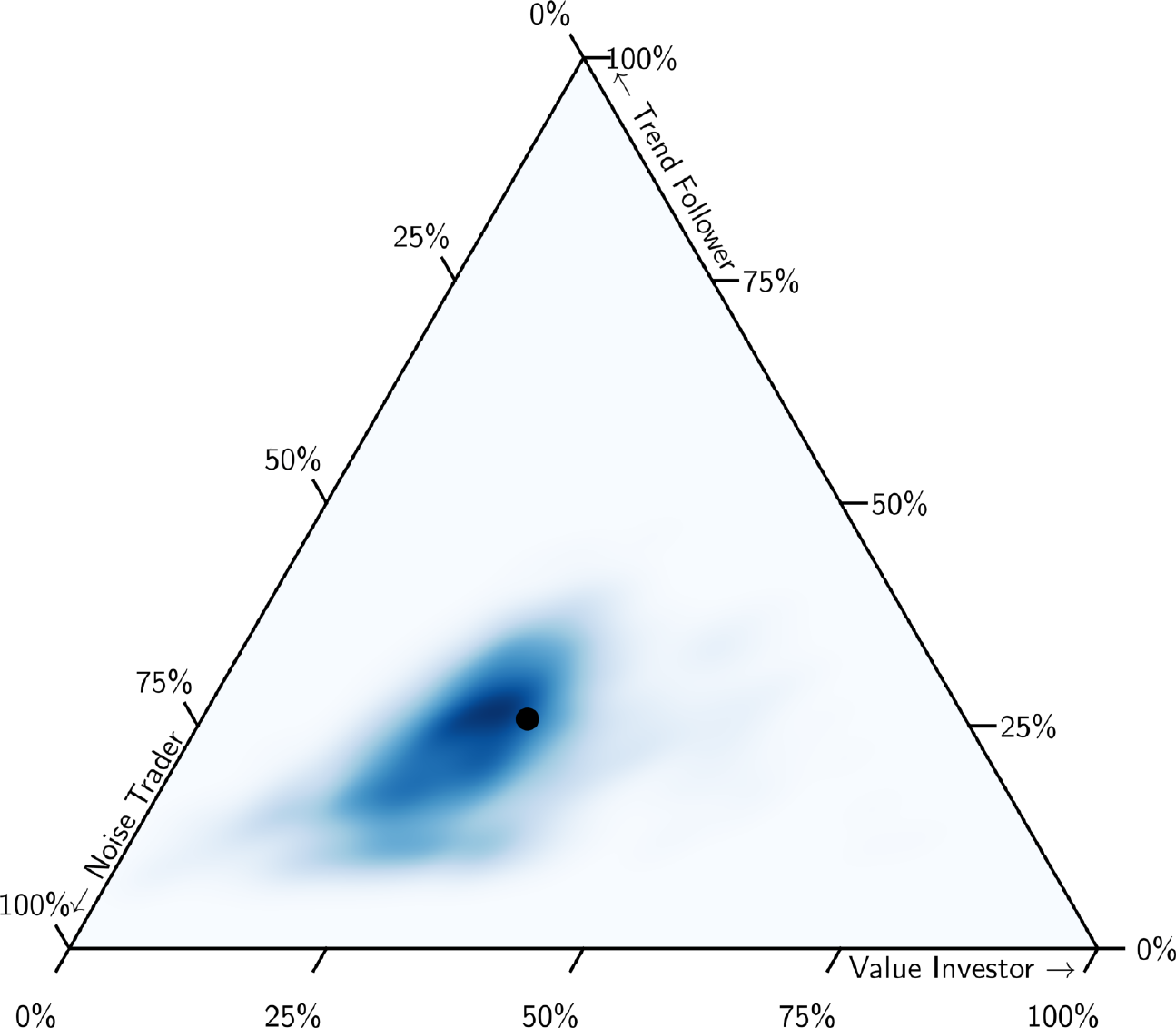}}\put(10,180){\Huge C}
        \end{subfigure}
        \hfill
        \begin{subfigure}   
            \centering  {\includegraphics[width=0.52\textwidth]{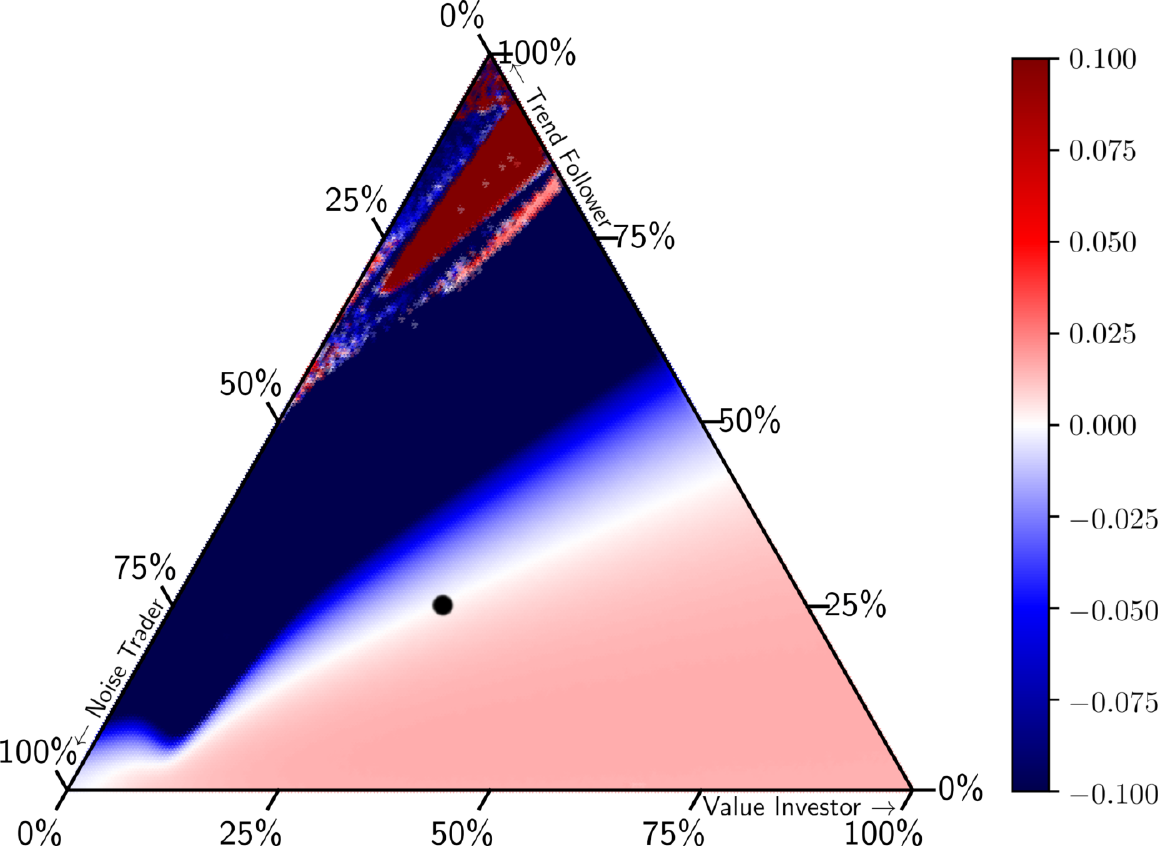}}\put(-10,190){\Huge D}
        \end{subfigure}
        \caption{{\bf Profit dynamics as a function of wealth.} {\bf A} shows how wealth evolves {\it on average} through time  under reinvestment. The intensity of the  color denotes the rate of change. {\bf B} shows sample trajectories for a few different initial values of the wealth vector, making it clear that the trajectories are extremely noisy due to statistical uncertainty, so that the deterministic dynamics of panel A is a poor approximation. The visualization displays three different initial wealth vectors, each color-coded. The marker \emph{+} indicates the initial wealth. The trajectories with the same color follow the system for T = 200 years and color saturation increases with time.  Starting from uniformly distributed initial conditions, {\bf C} displays a density map of the asymptotic wealth distribution after 200 years. The system is initialized at random with a uniformly distributed wealth vector and then allowed to freely evolve for 200 years. The color's darkness is proportional to density.   Panel {\bf D} displays the one time step autocorrelation of price returns. The black dot is the equilibrium point from Panel A.}
        \label{figure:efficiency_dynamics}
    \end{figure*}

We now investigate the dynamics of the market ecosystem.  To understand how the wealth of the strategies evolves through time, we allow reinvestment and plot trajectories corresponding to the average return at each wealth vector $\boldsymbol{w}$.  This is done by averaging over many different runs.  The result is shown in Figure~\ref{figure:efficiency_dynamics}A. Most of the wealth trajectories in the diagram evolve toward a fixed point where the wealths of the strategies no longer change.   There is also a region at the top of the diagram where the dynamics are more complicated (due to instabilities) and a region in the lower left corner where the ecology evolves toward the boundary of the simplex.

At the fixed point the annual returns to the three strategies are all equal to $\pi = 2.05\%$, which to an investor is statistically indistinguishable from the  $2\%$ return from simply buying and holding the stock.  We will loosely refer to this fixed point as the {\it efficient equilibrium}. We say ``loosely'' because the volatilities of the strategies are $4.07\%$ for noise traders, $6.76\%$ for value investors, $4.62\%$ for trend followers and $9.09\%$ for a buy and hold, so that with a more sophisticated model of fund flows the equilibrium might shift slightly in favor of strategies with less risk.  
A common way to measure the performance of a trading strategy is in terms of the ratio of the mean to the standard deviation of its returns, which is called the Sharpe ratio $\mathcal{S}$.  The corresponding Sharpe ratios (without subtracting the risk-free rate) are $0.50$, $0.31$ and $0.44$ for the three strategies, and $0.22$ for a buy and hold.  These Sharpe ratios and their variation are reasonable numbers for investment funds, indicating that our model ecosystem is roughly as efficient (or inefficient) as a real market (if anything the Sharpe ratios are a bit low).

The large central region of initial conditions that are attracted to the efficient equilibrium gives a misleading impression of a smooth evolution toward a state of market efficiency. In fact the dynamics are noisy and stray far from the deterministic dynamics shown in Figure~\ref{figure:efficiency_dynamics}A. Tracking a few individual trajectories, as we do in  Figure~\ref{figure:efficiency_dynamics}B, demonstrates that the dynamics are dominated by noise, due to the statistical uncertainty in the performance of the strategies. The typical trajectories bear little correspondence to the deterministic trajectories of Figure~\ref{figure:efficiency_dynamics}A, and the convergence to the efficient equilibrium is weak.

To get a feeling for the asymptotic wealth distribution, we sample the space of initial wealth uniformly, simulate the ecosystem dynamics under reinvestment with $f=1$, and record the final wealth after 200 years, as shown in Figure~\ref{figure:efficiency_dynamics}C.  The deviations from the efficient equilibrium point are substantial, often more than 20\%.  Furthermore, the evolution toward the asymptotic distribution is exceedingly slow:  Each trajectory in Figure~\ref{figure:efficiency_dynamics}B spans 200 years of simulated time.  There are substantial changes in the relative wealth taking place over time scales that are longer than a century. 

The long time scale to reach efficiency observed here matches with the estimate made by Farmer in reference \cite{Farmer2002MarketEvolution}.  In the ideal case of a stationary market and I.I.D. normally distributed returns, the time required to detect excess performance $\Delta \mathcal{S}$ with a statistical significance of $s$ standard deviations is approximately $\tau = (s/\Delta \mathcal{S})^2$. To take an example, a buy and hold of the S\&P index has a Sharpe ratio of roughly $\mathcal{S} = 0.5$.  
For a strategy whose annualized Sharpe ratio is superior by $\Delta S = 0.1$, a $20\%$ improvement over a buy and hold, roughly $400$ years are required to confirm its superior performance with two standard deviations.  Furthermore, as shown in reference \cite{Cherkashin2009TheGameb}, the approach to market efficiency follows a power law of the form $t^{-\alpha}$, where $0 \le \alpha \le 1$.  For large times this is much slower than an exponential. This happens because the approach to efficiency slows down as the market becomes more efficient. Near the efficient equilibrium the dynamics are dominated by the noise.

To demonstrate that statistical uncertainty is the dominant factor determining the approach to efficiency we did a series of experiments that are described in detail in the supplementary appendix (section ).  As the reinvestment rate $f$ in Eq.~(\ref{wealth_reallocation}) varies in the range $0.1 \le f \le 3$  the rate of approach to the asymptotic wealth distribution remains roughly the same.  In contrast, varying the noise trader volatility, which affects the statistical uncertainty in the performance of all three strategies, has a substantial effect.

The absence of autocorrelation in price returns is an indicator of market efficiency.  Efficient price returns should have an autocorrelation that is reasonably close to zero (close enough that it is not possible to make statistically significant excess profits).  In Figure~\ref{figure:efficiency_dynamics}D we plot the one time step autocorrelation of returns across the wealth landscape.  There is a striking white band across the center of the simplex, corresponding to zero autocorrelation.  This happens when trend followers invest about 40\% of the total wealth, thereby eliminating the autocorrelation coming from the dividend process.  The wealth of trend followers fluctuates, even at very long times, and consequently the degree of autocorrelation in price returns fluctuates as well.

\subsection{Community matrix}

 The community matrix is a tool used in ecology to describe the pairwise effects of the population of species $j$ on the population growth rate of species $i$ \cite{May1972WillStable,Novak2016CharacterizingMatrix}.  As originally pointed out by Farmer \cite{Farmer2002MarketEvolution}, who called it the gain matrix, an analogous quantity is also useful for interpreting the behavior of market ecosystems.  
 Let $\pi_i$ be the average return of strategy $i$, i.e. $\pi_i = \lim_{T \to \infty} \left( \prod_{t=1}^T \pi_i(t) - 1 \right)^{1/T}$.  The analogue of the community matrix for market ecology is
     \begin{equation}\label{eqn:community}
         G_{ij} = \frac{\partial \pi_i}{\partial w_j}.
     \end{equation}
This has units of one over time.  The wealth $w_i(t)$ invested in strategy $i$ replaces the population size of a species.  The possible pairwise interactions between strategies can be classified according to the sign of $G_{ij}$.  If both $G_{ij}$ and $G_{ji}$ are negative, then strategies $i$ and $j$ are competitive; if $G_{ij}$ is positive and $G_{ji}$ is negative, then there is a predator-prey interaction, with $i$ the predator and $j$ the prey; and if both $G_{ij}$ and $G_{ji}$ are positive, then there is a mutualistic interaction \cite{Bronstein1994ConditionalInteractions}.

Because we do not have a differentiable model for our toy market ecology, we compute the community matrix numerically using finite differences (see Materials and Methods).  The community matrix is strongly density dependent. If we compute the community matrix near the equilibrium point in the center of the simplex, we get the result shown in Table 1. 

The diagonal entries are all negative, indicating that the strategies are competitive with themselves.  This means that their average returns diminish as the strategy gets larger, causing what is called {\it crowding} in financial markets.  We already observed this in Figure~\ref{figure:performance_dd_sharpe}. Interestingly, however, the size of the diagonal terms varies considerably, from $-0.89$ for noise traders to $-19.3$ for trend followers.  This means that we should expect trend followers to experience crowding more strongly than noise traders.

All the other entries are positive, indicating mutualism.  This implies that every strategy benefits from an increase in the wealth of any of the other strategies.  While we initially found it surprising that all the strategies could have mutualistic interactions with each other, on reflection this makes sense:  the ecology is by definition efficient at the equilibrium, and driving any of the strategies away from equilibrium creates an inefficiency that provides a profit opportunity for the other two strategies.  (It is not clear to us whether this is specific to this particular set of strategies, or whether a richer set of strategies would display more complicated behavior near an efficient equilibrium).

The community matrix is density dependent.  If we compute the community matrix at the wealth vector given in Table~\ref{tab:num_interaction2}, where the value investors are dominant, there is a shift in the pairwise community relations.  
\begin{table}
        \centering
        \caption{ Estimated community matrix near the equilibrium at $\boldsymbol{w} = (NT = 0.43, VI = 0.34, TF = 0.23)$.
                }
        \label{tab:num_interaction1}
        \begin{tabular}{l|rrr}
        \toprule
         ${G}_{ij}$ &        NT &        VI &        TF \\
        \midrule
            NT &  -0.89\% &   0.89\% &   0.82\% \\
            VI &  26.6\% & -10.6\% &  22.4\% \\
            TF &  11.1\% &  15.2\% & -19.3\% \\
        \bottomrule
        \end{tabular}
        \caption{{\bf Estimated community matrix near} $\boldsymbol{w} = (NT = 0.26, VI = 0.55, TF = 0.19)$.}
        \label{tab:num_interaction2}
        \begin{tabular}{l|rrr}
        \toprule
         ${G}_{ij}$ &        NT &        VI &        TF \\
        \midrule
            NT & -0.46\% &  0.40\% &  0.36\% \\
            VI &  8.94\% & -0.77\% & -1.89\% \\
            TF &  6.81\% &  6.87\% & -9.65\% \\
        \bottomrule
        \end{tabular}
    \end{table}
As before, all of the terms in the row corresponding to the noise traders are small, indicating that the noise traders are not strongly affected by other strategies, and that they compete only weakly with themselves.  This should not be surprising -- the noise traders' strategy is mostly random, and is less influenced by prices than the other two strategies.  Value investors, who have the majority of the wealth in this case, still strongly benefit from an increase in the wealth of noise traders (though less so than at the equilibrium).   However, there is now a second negative term in the second row, corresponding to the effect of trend followers.  In contrast, from the third row we see that trend followers benefit from an increase in the wealth of both noise traders and value investors, implying that trend followers now prey on value investors.   Other variations in community relationships can be found at different points in the wealth landscape, illustrating density dependence.   

The Lotka-Volterra equations, which describe how the populations in an idealized predator-prey system evolve through time, are perhaps the most famous equations in population biology.  Their surprising result is that, at some parameter values, they have solutions that oscillate indefinitely.  Using the assumption of no density dependence, Farmer derived Lotka-Volterra equations for market ecology \cite{Farmer2002MarketEvolution}.  Our results here indicate that the density dependence in this system is so strong that simple Lotka-Volterra equations are a poor approximation, at least for this system.  The existence of oscillating solutions in financial markets remains an open question. 

\subsection{Food Webs and Trophic Level}\label{section:foodweb}
    
The food web provides an important conceptual framework for understanding the interactions between species.  If lions eat zebras and zebras eat grass, then the population of lions is strongly affected by the density of grass, and similarly the density of grass depends on the population of lions, even though lions have no direct interactions with grass.  The trophic level of a species is by definition one level higher than what it eats, so in this idealized system grass has trophic level one, zebras have trophic level two and lions trophic level three.  

The existence of animals with more complicated diets, such as omnivores and detritivores, means that real food webs are never this simple.  If we let $A_{ij}$ be the share of species $j$ in the diet of species $i$, then the trophic level $T_i$ of species $i$ can be computed by the relation
    \begin{equation}\label{eqn:trophiclevel}
        {T}_i = 1 + \sum_{j} A_{ij}  {T}_j. 
    \end{equation}
The resulting trophic levels are typically not integers, but they still provide a useful way to think about the role that a given species plays in the ecology.

We can also compute trophic levels for the strategies in a market ecology.  We define the analogous quantity $A_{ij}$ as the fraction of the returns of strategy $i$ that can be attributed to the presence of strategy $j$.  We do this by simply comparing the returns of strategy $i$ at wealth $\boldsymbol{W}$ to those when strategy $j$ is removed, i.e. when $W_j = 0$ but all the other wealths remain the same.  In mathematical terms,
\begin{equation}
    A_{ij} = \max\left[0, \pi_i(W_1, \ldots, W_j, \ldots, W_N) - \pi_i(W_1, \ldots, 0, \ldots, W_N)\right].
\end{equation}
The maximum is taken so that $A_{ij}$ is never negative.  For computing the trophic levels we only care about the strategies that $i$ benefits from, not those that cause it losses.

Equations (9) and (10) allow us to compute trophic levels for each of the strategies. At the equilibrium point, for example, the trophic levels are (Noise Trader = 1, Value Investor = 2, Trend Follower = 3)
In order to better understand the density dependence, we compute trophic levels at each point in the wealth landscape.  For three strategies there are $3! = 6$ possible orderings of the trophic levels.  We display the ordering of the trophic levels across the wealth landscape in Figure~\ref{figure:foodweb_ordering}.

\begin{figure}[htb]
    \centering\includegraphics[width=\linewidth]{trophic_web_regions_filter16_printer.pdf}
    \caption{{\bf A survey of the trophic levels across the wealth landscape.} We color the diagram according to the increasing ordering of the trophic levels of the three strategies (see legend).  Red, for example, denotes the dominant zone where the noise traders have the lowest trophic level and trend followers have the highest trophic level (with value investors in the middle).  In the grey region there are cycles where the trophic levels become undefined. The black dots correspond to samples of the wealth vector after 200 years, as shown in Figure 3(c). The system spends most of its time in the grey and red zones.}\label{figure:foodweb_ordering}
\end{figure}

The computation of trophic levels is complicated by the fact that, for some wealth vectors, there are cycles in the food web.  For example, for $\boldsymbol w = (0.05, 0.15, 0.80)$, value investors exploit noise traders (who cause reversion to fundamental value), trend followers exploit value investors (who induce autocorrelations), and the noise traders exploit trend followers (who generate excess volatility), to complete a cycle.   When this happens equation (9) does not converge and the trophic levels become undefined.  Cycles in the trophic web are not unique to markets -- they can also occur in biology, for example due to cannibalism or detritovores.

A comparison of Figure 4 to Figure 3C makes it clear that at long times (after transients have died out), the system divides its time between the region in which the trophic levels are ordered as (Noise Trader, Value Investor, Trend Follower), as they are at the equilibrium point, and the region where there are cycles, where the trophic levels are undefined.


\subsection{How ecosystem dynamics cause market malfunction}\label{section:malfunction}
    \begin{figure}[b!]
    \vspace{-2em}\begin{subfigure}
        \centering
        \put(0,0){\includegraphics[width=\linewidth]{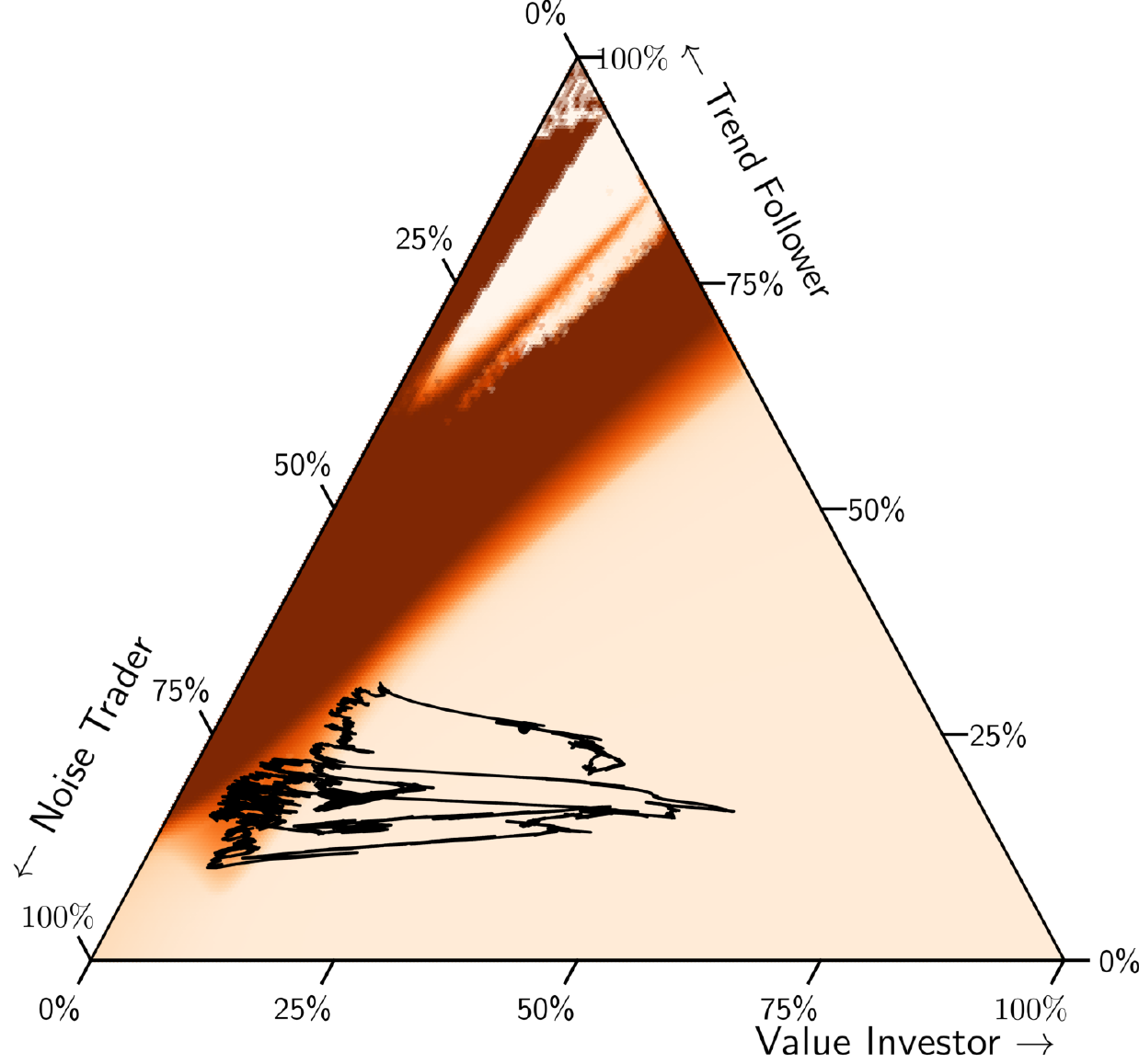}}\put(200,190){\sffamily\huge A}\\
    \end{subfigure}
    \begin{subfigure} 
        \centering
        \includegraphics[width=\linewidth]{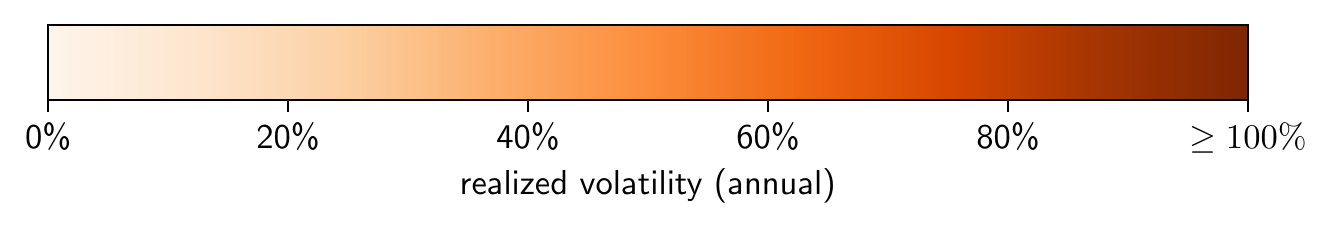}
    \end{subfigure}
    \begin{subfigure}  
        \centering
        \includegraphics[width=\linewidth]{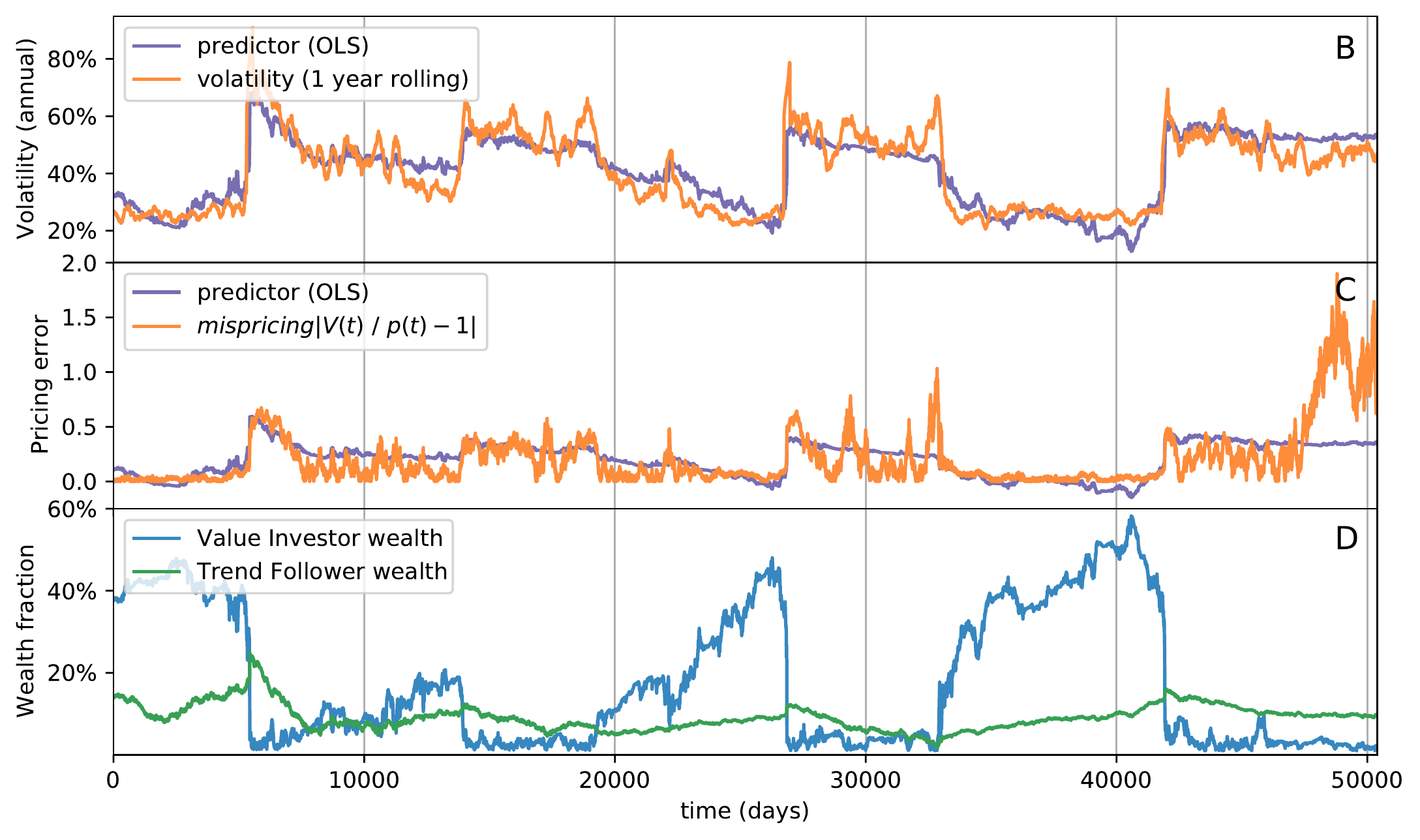}
    \end{subfigure}
        \caption{{\bf How ecosystem dynamics cause market malfunctions.}  
        Panel {\bf A} gives a color map of the price volatility over the wealth landscape; the volatility is low and constant throughout  the lower right part of the diagram, where the system spends most of its time, but there is a high volatility region running across the upper left.  A sample trajectory spanning 200 years beginning at the efficient equilibrium is shown in black. The noise caused by statistical fluctuations in performance causes large deviations from equilibrium and excursions into the high volatility region. Panel {\bf B} shows the volatility of this trajectory as a function of time, plotted against the predicted volatility (see Eq.~(\ref{eqn:predictors})). 
        Panel {\bf C} shows the actual mispricing plotted against the predicted mispricing. 
        Panel {\bf D} shows the wealth of the value investors and trend followers.    
        }
        \label{figure:volatility_mispricing}
    \end{figure}
    
The wealth dynamics of the market ecosystem help explain why the market malfunctions and illuminate the origins of excess volatility and {\it mispricing}, i.e. deviations of prices from fundamental values.  Both in real markets \cite{Shiller1989} and in the agent-based models mentioned earlier (including our model here), volatility and mispricings change endogenously with time -- there are eras where they are large and eras where they are small.   Volatility varies intermittently, with periods of low volatility punctuated by bursts of high volatility -- this is called clustered volatility.   The standard explanations for clustered volatility are fluctuating agent populations \cite{Arthur1997AssetMarket, brock1997,Franke2012} and leverage \cite{Thurner2011LeverageVolatility}.  We focus here on the first explanation; while we also observe that clustered volatility increases with increasing leverage, we have not investigated this in detail here.

Figure~\ref{figure:volatility_mispricing}A presents the variation of the volatility across the wealth landscape. The landscape can roughly be divided into two regions.  On the lower right there is a flat low volatility ``plain''  occupying most of the landscape.  On the upper left there is a high volatility region, with a sharp boundary between the two.  As we will now show, excursions into the high volatility region cause clustered volatility.  A similar story holds for mispricing.   

Figure~\ref{figure:volatility_mispricing}A shows a sample trajectory that begins at the efficient equilibrium and spans 200 years. This sample is representative of trajectories that fluctuate around the equilibrium point indefinitely. We choose a 200-year time span to show the scale at which spikes in volatility and mispricing occur. The statistical fluctuations in the performance of the three strategies act as noise, causing large excursions away from equilibrium.  The trajectory mostly remains on the volatility plain, but there are several epochs where it ventures into the high  volatility region, causing bursts of high volatility.

\begin{table}[tbhp!]
    \centering
    \caption{Multivariate regressions with volatility and mispricing as dependent variables, and the funds' wealth as independent variables. }\label{table:regression}
    \begin{tabular}{lll}
    {\bf volatility} & $R^2 = 0.79$ & observations: 50,397\\ \toprule
    
    independent variable & coefficient & t \\ \midrule
    noise trader & 2.4 &  10 \\
    value investor & -68 &  -249 \\
    trend follower & 107 &  169 \\
    \\
    {\bf mispricing} & $R^2 = 0.33$ & observations: 50,397\\ \toprule
    independent variable & coefficient & t \\ \midrule
    noise trader & -0.15 &  -18 \\
    value investor & -1.02 &  -107 \\
    trend follower & 1.5 &  69 \\
    \bottomrule
    \end{tabular}
    \ifdefined\longformpaper
    \fi
    \end{table}

The wealth dynamics have strong explanatory power for both mispricing and volatility.   This is illustrated in Table~\ref{table:regression}, where we perform regressions of the strategies' wealth against volatility using daily values for the time series shown in Figure~\ref{figure:volatility_mispricing}A, for volatility $R^2 = 0.79$ and for mispricing $R^2 = 0.33$.  In both cases the value investor's wealth and the trend follower's wealth have large coefficients (in absolute value), and the fit is overwhelmingly statistically significant.  The noise trader is also highly statistically significant, but the coefficients and the t-statistics are more than an order of magnitude smaller. In Figures~\ref{figure:volatility_mispricing}B and ~\ref{figure:volatility_mispricing}C we compare a time series of the predicted volatility and predicted mispricing against the actual values. The predictions are very good.
\begin{align}
\begin{split}
    \hat{\nu} & = -68 w_{vi} + 107 w_{tf} + 2.4 w_{nt}\\
    \hat{m} & = -1.02 w_{vi} + 1.5 w_{tf} -0.15 w_{nt} \\ 
\end{split}
\label{eqn:predictors}
\end{align}
Note that in both cases the coefficient for trend followers is positive, indicating that they drive instabilities, and the coefficient for value investors is negative, indicating their stabilizing influence.  

Nonetheless, due to their effect on the population of value investors, the net effect of the trend followers on market malfunctions is not obvious.   In Figure~\ref{figure:volatility_mispricing}D we plot the wealth of value investors and trend followers. The strong mutualism predicted by the community matrix is clearly evident from the fact that the wealths of trend followers and value investors rise and falls together.  However, their dynamics are quite different -- there are several precipitous drops in the value investors' wealth, whereas the trend followers tend to take more gradual losses.  As predicted, the highest volatility episodes happen when the value investors' wealth drops sharply while the trend followers'  wealth is high.

\section*{Discussion}

Our analysis here demonstrates how understanding fluctuations of the wealth of the strategies in the ecology can help us predict market malfunctions such as mispricings and endogenously generated clustered volatility.
The toy model that we study here is simple and highly stylized, but it illustrates how one can import ideas from ecology to better understand financial markets.  Our analysis of this model illustrates several properties of market ecosystems that we hypothesize are likely to be true in more general settings.  

Concepts from ecology give important insights into how deviations from market efficiency occur and how they affect prices. While the market may be close to efficiency in the sense that the excess returns to any given strategy are small, there can nonetheless be substantial deviations in the wealth of different strategies, that can cause excess volatility and market instability.

Market ecology is a complement rather than a substitute for the theory of market efficiency.  There are circumstances, such as pricing options, where market efficiency is a useful hypothesis.  Market ecology, in contrast, provides insight into how and why markets deviate from efficiency, and what the consequences of this are.  It can be used to explain the time dependence in the returns of trading strategies, and in some cases it can be used to explain market malfunctions.
One of our main innovations here is to demonstrate how to compute the community matrix and the trophic web, which provide insight into the interactions of strategies.

There are so far only a few examples of empirical studies of market ecology \cite{KIRILENKO2017, Musciotto2018}.  This is because such a study requires counterparty identifiers on transactions in order to know who traded with whom.  Trying to study a market ecology without such data is like trying to study a biological ecology in which one can observe that an animal ate another animal without any information about the types of animals involved.    Unfortunately, for markets such data is difficult for most researchers to obtain.  

Regulators potentially have access to the balance sheets of all market participants, which can allow them to track the ecology of the markets they regulate in detail.  Ideas such as those presented here could provide valuable insight into when markets are in danger of failure, and make it possible to construct models for the ecological effect of innovations, e.g. the introduction of new types of assets such as mortgage-backed securities.

One of our most striking results is that the approach to efficiency is highly uncertain and exceedingly slow. As already pointed out, this should be obvious from a straightforward statistical analysis, but it is not widely appreciated.  Our results demonstrate this dramatically and they indicate that, even in the long-term, we should expect large deviations from efficiency. 

There are many possible extensions to this work.  An obvious follow up is to explore a larger space of strategies, or to let new strategies evolve in an open-ended way through time.  Does the process of strategy innovation tend to stabilize or destabilize markets?  Another follow up is to construct a model that is empirically validated against data with counterparty identifiers.  Our analysis here provides concepts and methods that could be used to interpret the behavior of real world examples.
\matmethods{
{\bf Accounting and Balance Sheets. } 
    The funds in our model use a stylized balance sheet that is presented in table \ref{table:balancesheet}. Funds are endowed with equity capital $K = W(0)$, in the form of cash $C$ in dollars and shares of trading securities $S$. When $S > 0$, the fund holds this amount of securities, and when $S < 0$, it has borrowed this amount from other market participants, to create a short position. In order to guarantee that the short-selling fund can return the borrowed securities to the lender at a later time, the fund sets aside a margin amount $M$ equal to the current market value of the borrowings, in the form of cash. The funds can use \emph{leverage}, meaning using borrowed funds to purchase additional risky assets, by borrowing cash $L$. Public regulatory filings of U.S. institutional fund managers indicate common leverage ratios between 1 and 10\footnote{\emph{The Electronic Data Gathering, Analysis, and Retrieval system (EDGAR) \url{https://www.sec.gov/edgar.shtml}}. Securities and Exchange Commission}.  The interest rate that  applies to cash holdings, loans, and margin is the same as the risk-free rate from holding the bond.
    
    \begin{table}
        \centering
        \caption[Balance Sheets]{This table details the balance sheet items used by all funds. All securities use the most recent market value in valuation. }
        \label{table:balancesheet}
        \begin{tabular}{rl|rl}
        \textbf{Assets} &  & \textbf{Liabilities} &  \\ \hline\hline
        & & \textbf{Equity} & \\ \cline{3-4}
        cash & $C$ & capital & K\\
        & & \textbf{Debt}  & \\ \cline{3-4} 
        margin & $M$ & loans & $L$ \\
        trading securities & $S^+$ & borrowed securities & $S^-$ \\
        \end{tabular}
    \end{table}    
    
    Wealth is calculated as:
        
    \begin{equation}
    \label{equation:workingcapital}
        W(t)
        = C(t) + S(t)p(t) - L(t),
    \end{equation}
    and changes according to
    \begin{equation}\label{equation:wealth_evolution}
        W(t{+}1) = W(t) + f \left[r \left(C(t){-}L(t)\right){+} \left(p(t{+}1){-}p(t){+} D(t)\right)S(t) \right].
    \end{equation}
    
    A fund can only violate its leverage constraint when the proportion of risky assets changes faster than the amount of equity capital. This can happen due to losses. 
    We require that all funds meet the solvency condition $ W(t) > 0$. The simulation ends when one or more funds are insolvent. 
    Table~\ref{table:default_params} lists the default parameters.
    
    \begin{table}
    \ifdefined\longformpaper
    \fi
    \centering
    \caption{This table provides a listing of the model parameters and their values. }\label{table:default_params}
    \begin{tabular}{lll}
    parameter & value & description \\
    \midrule
    $r$ & $1\%$ annual &  risk-free rate  \\
    $g$ & $1\%$ annual &  dividend growth rate  \\
    $k$ & $2\%$ annual &  cost of equity \\
    $\sigma$ & $6\% $ annual & dividend growth volatility  \\
    $\omega$ & $0.1$ & dividend autocorrelation parameter \\
    $f$ & 1 & reinvestment rate  \\
    $\rho$ & $1-\sqrt[6\cdot 252]{\frac{1}{2}}$  & noise trader mean reversion rate \\
    $\sigma^{\mathtt{NT}}$ & $12\%$ annual & noise trader volatility \\
        $\lambda^*_{\mathtt{NT}}, \lambda^*_{\mathtt{VI}}, \lambda^*_{\mathtt{TF}}$ & 1,8,1 & leverage limit \\
    $c_{\mathtt{NT}}, c_{\mathtt{VI}}, c_{\mathtt{TF}}$ & 5, 10, 4 & signal scale\\
    \bottomrule
    \end{tabular}
    \ifdefined\longformpaper
    \fi
    \end{table}

    {\bf Model and Software. } The simulation in this paper builds on the \emph{Economic Simulation Library}, an open-source library for agent-based modeling which is accessible at~\url{https://github.com/INET-Complexity/ESL}. The model and code to run the experiments in this paper is available at~\url{https://github.com/INET-Complexity/market-ecology}.
}

\showmatmethods{}
\acknow{We thank Klaus Schenk-Hopp\'{e}, Robert MacKay, Michael Wooldridge, Rama Cont, Alissa Kleinnijenhuis and Cephas Svosve for enlightening discussion during the development of this work.
We acknowledge funding by J.P. Morgan AI Faculty Awards, Baillie-Gifford and the Rebuilding Macroeconomics program, funded by the Economic and Social Research Council. This research was partially supported by TAILOR, a project funded by EU Horizon 2020 research and innovation programme under GA No 952215.
}
\showacknow{}  
\bibliography{references}


\ifdefined\longformpaper

\pagebreak[5]

\appendix 

\section*{Supplementary Materials}

\section{Statistical Properties of the Model}

\subsection*{Dividends}
The dividend process has the following properties. Consider the auxiliary autocorrelated process $U(t)$ from equation \ref{label:multivariate_geometric_brownian_motion_autocorrelation}, which is a modified \texttt{AR} $(1)$ model.

\begin{equation}
    \begin{split}
         \mathtt{Var}(U(t)) &= \mathtt{Var}(\omega U(t-2) + (1-\omega^2)Z(t)) \\&= \omega^2 \mathtt{Var}(U(t-2)) + \mathtt{Var}((1-\omega^2)Z(t)) \\
         &=1\\
    \end{split}
\end{equation}

Therefore the dividend process underlying the autocorrelated process $U(t)$ has variance 1. This is useful in the construction of the dividend process, as from this it follows that the dividend process' log returns have a variance of $\sigma$.

We now investigate the autocorrelation of the process $U(t)$. We proceed by induction, and it is useful to know that in the base case, $\mathtt{Cov}(U(t), U(t)) = 1$. The lag $\tau$ autocorrelation function of the process $U(t)$ is given by

\begin{equation}
    \begin{split}
        \mathtt{Cov}(U(t),& U(t{-}\tau))\\ &= \mathtt{Cov}(\omega U(t{-}2){+}(1{-}\omega^2)Z(t),U(t{-}\tau))\\
        &= \omega\mathtt{Cov}(U(t{-}2),U(t{-}\tau)){+}(1{-}\omega^2)\mathtt{Cov}(Z(t), U(t{-}\tau))\\
        &= \omega \mathtt{Cov}(U(t{-}2), U(t{-}\tau))\\ &= \begin{cases} \omega^{\tau-1} & \tau > 1 \\ 0 & \tau = 1 \\ \end{cases}    \\
    \end{split}
\end{equation}

\subsection*{Market Prices}

We illustrate aspects of market prices generated by the model in figure \ref{fig:stylized_facts}.

\begin{figure}[htb]
    \centering
    \includegraphics[width=\linewidth]{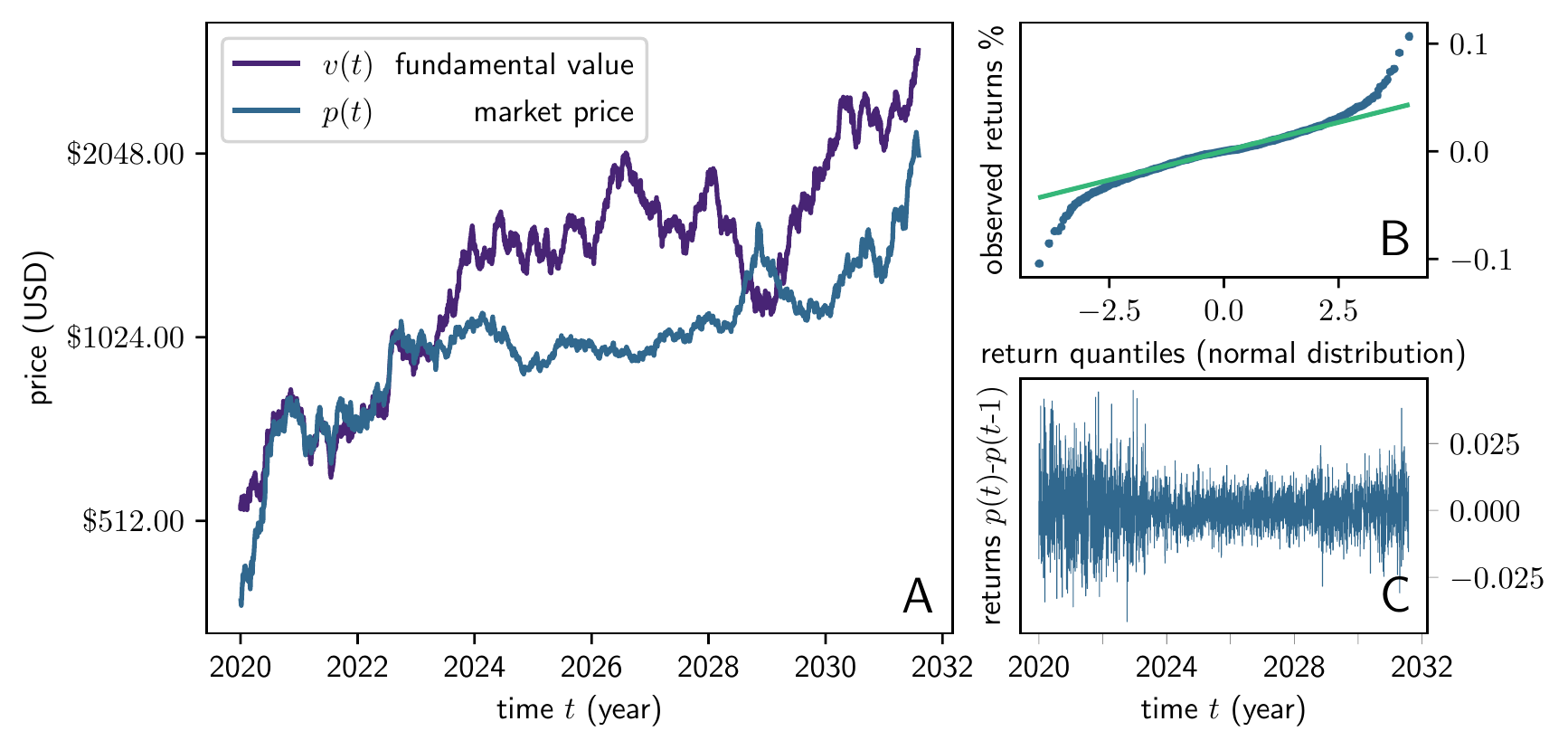}
    \caption{A brief summary of aspects of market prices. Panel A: market prices track fundamental over long periods, are often within a factor of two, yet are noisy. Panel B: the returns distribution of market prices has heavy tails, as is evidenced by the deviation of the empirical distribution (markers) from the return quantiles expected by a process with normally distributed returns (line). Panel C: market prices display cluster volatility, with large price moves likely followed by large moves resulting in clustering in time. }
    \label{fig:stylized_facts}
\end{figure}

\section{ Market Clearing } 
    Prices are set by a price setter who chooses prices such that demand and supply match as close as possible. The excess demand of agent $a$ for the stock is defined in equation~\ref{eqn:excess_demand}. The market excess demand curve for the stock is the aggregate of the excess demand of all agents $a$. As in the classical Walrasian setting, the price setter seeks to match demand and supply, so that aggregate excess demand is zero for each investment, by finding a root of the market excess demand curve. 
    
    \begin{equation}
        \sum_{a\in\mathcal{A}} E_a(p) = 0 
    \end{equation}
    
    However, if no solution is found through the root-finding process, we must fall back to a heuristic that seeks for the best solution that only partially clears the market. We interpret the goodness of a solution as the extent to which the solution minimizes demand and supply \emph{mismatch}. We here use the square of excess demand, and this way the root-finding problem is transformed into the corresponding minimization problem:
    \begin{equation}
    \begin{aligned}
    & \underset{p}{\text{minimize}}
    & & \left( \sum_{a \in \mathcal{A}} E_{a}({p})\right)^2\\
    & \text{subject to}
    & & \text{finite~}p.
    \end{aligned}\label{eqn:aggregate_excess_demand_min} \\
    \end{equation}
    
    In practice, we rarely encounter situations where no clearing price is found, as they only occur in the chaotic regions where trend followers have amassed most wealth, and the system does not often visit that region.
    For univariate root-finding problems, methods such as Steffensen's method \cite{Johnson1968OnMethod} can quickly find a root of a continuous function using the derivative of the function. For the types of demand and supply functions we permit, this process is not guaranteed to find the nearest market-clearing price, nor is it guaranteed to find a clearing price at all, since there is no guarantee of existence of such a solution. If the root-finding procedure fails, we fall back on minimizing the clearing error (equation~\ref{eqn:aggregate_excess_demand_min}). In the spirit of Walras' auction, we use quasi-Newton methods to gradually move into the direction that decreases the clearing error fastest. Limited-memory BFGS(L-BFGS) is chosen as the minimization algorithm. As with many non-convex optimization problems, convergence is not guaranteed, though the chosen method compares favorably to others \cite{Cooke2012ImperialMarkets}. The method requires at each iteration the gradient in numerical form. The starting point for the minimization algorithm is the market-clearing prices for the previous time point.  Quasi-Newton methods that are widely used in numerical optimization rely on reconstructing a gradient of the function, and an analytic expression for the gradient is often not available. If the method uses a numerical scheme to reconstruct the gradient, such as the central difference method, this requires multiple evaluations of the function, and these schemes are prone to estimation errors. Automatic differentiation is used to obtain both the value, gradient and Jacobian in as few evaluations of the function as possible. We use the software library presented in \cite{Hogan2014FastC++}, which uses C++ meta-programming rules to rewrite equation~\ref{eqn:excess_demand} to compute not only the excess demand but also the gradient and Jacobian. With an eye towards future work, we develop a general solution that can deal with multiple assets.  

\section{Long-term dynamics}

We investigate the distribution of system system states $\boldsymbol{W}(t)$ at time $t$, as $t\to\infty$. 
Let $\mathcal{P}(t)$ and $\mathcal{W}(t)$ be discrete probability distributions defined on the space $\boldsymbol{W}$. $\mathcal{W}$ will be  the distribution of the system states of the model as it is presented in the main text. $\mathcal{P}(t)$ will be the distribution under a change of model parameters, defined by a series of experiments that are defined later.

Probability distributions such as $\mathcal{W}$ will estimated using N Monte-Carlo samples

$$
  \mathcal{W}_t \approx \frac{1}{N} \sum_{w\in \boldsymbol{W}} 1_{W(t) = w}
$$

A measure to describe the difference between the asymptotic distribution $\mathcal{W}_\infty$ and the distribution over states under an altered parameter choice $\mathcal{P}$ is the Kullback-Leibler divergence, in this case given by $$D_{\mathtt{KL}}(\mathcal{W}_\infty||P) = -\sum_{w\in\boldsymbol{W}} \mathcal{W}_\infty(w) \log \left( \frac{\mathcal{P}(w)}{\mathcal{W}_\infty(w)}\right) $$

The properties of this measure is that identical distributions have a distance of $D_{\mathtt{KL}} = 0$, and that the measure is unbounded; at our initial condition we find that $D_{\mathtt{KL}}\approx 0.98$.

\subsection*{Numerical Computation}

Assume observations $w_1$ and $w_2$ are distributed as multivariate Gaussians $\mathcal{N}_1 = (\mu_1, \Sigma_1)$ and $\mathcal{N}_2 = (\mu_2, \Sigma_2)$

\begin{multline}
D_{\mathtt{KL}} (\mathcal{N}_1, \mathcal{N}_2) = \frac{1}{2} \biggl[ \mathtt{tr}(\Sigma_2^{-1}\Sigma_1) \\ + (\mu_2 - \mu_1)^T \Sigma^{-1}_2 (\mu_2 - \mu_1)\\ - k  + \ln \left(\frac{|\Sigma_2|}{|\Sigma_1|} \right) \biggr] 
\end{multline}

where $k$ is the number of dimensions ($k=3$ in the case of three investment strategies).

\subsection*{Reinvestment Rate}

In the literature, a positive relationship between fund performance and subsequent investment by external investors is found \cite{Chevalier1997RiskIncentives}. This relationship however is non-linear but convex, meaning that the higher the funds performance, the larger the inflow of new capital over the next period. There are other factors that play a role, such as that new funds have different investor dynamics than more established funds. 

We simplify this complex relationship by defining $f$ to be the reinvestment rate, meaning the rate at which external investors to the fund invest additional capital when the fund performs well, and withdraw when the fund under-performs the benchmark rate of return $r$: $W(t+1) = W(t) (1 + f \pi(t)).$

This is achieved by depositing additional cash into the fund when $\pi(t) > 0$, and decreasing cash and stock when $\pi(t) < 0$.

The study \cite{Chevalier1997RiskIncentives} shows that for established funds, when a fund earns an annual return of 25\%, the fund sees an additional inflow of 50\%, corresponding with $f=3$. However, this is an estimate of the highest reinvestment ratio, as for most cases the inflows and outflows are much smaller, as the authors showed the relationship to be convex in fund performance.

\subsection*{Determining the time to convergence}

We approximate the asymptotic distribution by taking the average of multiple samples of long runs of the simulation. For this we choose a horizon of $T=200$ years. We choose initial conditions different from the observed equilibrium state, so that we can tell apart transient states and stochastic steady states more clearly. The initial conditions are endowments distributed uniform in the strategy simplex. 

The chosen horizon of $200$ years is sufficient to bring down the distance by four fifths, after which random fluctuations dominate.

\subsection*{Results: Reinvestment Rate}\label{section:reinvestment}

We investigate how the reinvestment rate influences the time to converge to the steady state. For this, we vary the reinvestment rate $f$ ranging from the values $f=0.1, 0.3, 1.0, 3.0$ to $5.0$. The first thing that stands out is that when the reinvestment rate is $5$, the systems does not converge. Inspection reveals that this happens because for this parameter choice, the system almost always ends up with one of the funds in default, meaning the system moves to a point on the boundaries of the simplex far away from the equilibrium point.
For smaller choices of $f$, we observe eventual convergence. Because the reinvestment rate affects the equilibrium state, we only measure the Kullback-Leibler divergence to the long-term $t=100Y..200Y$ distribution of trajectories with the same reinvestment rate.
However, for each test we initialize the system to the same uniform distribution. The uniform distribution will have a different KL-divergence value to the different equilibrium states for each choice of $f$.

\begin{figure}[htb]
    \includegraphics[width=1.0\linewidth]{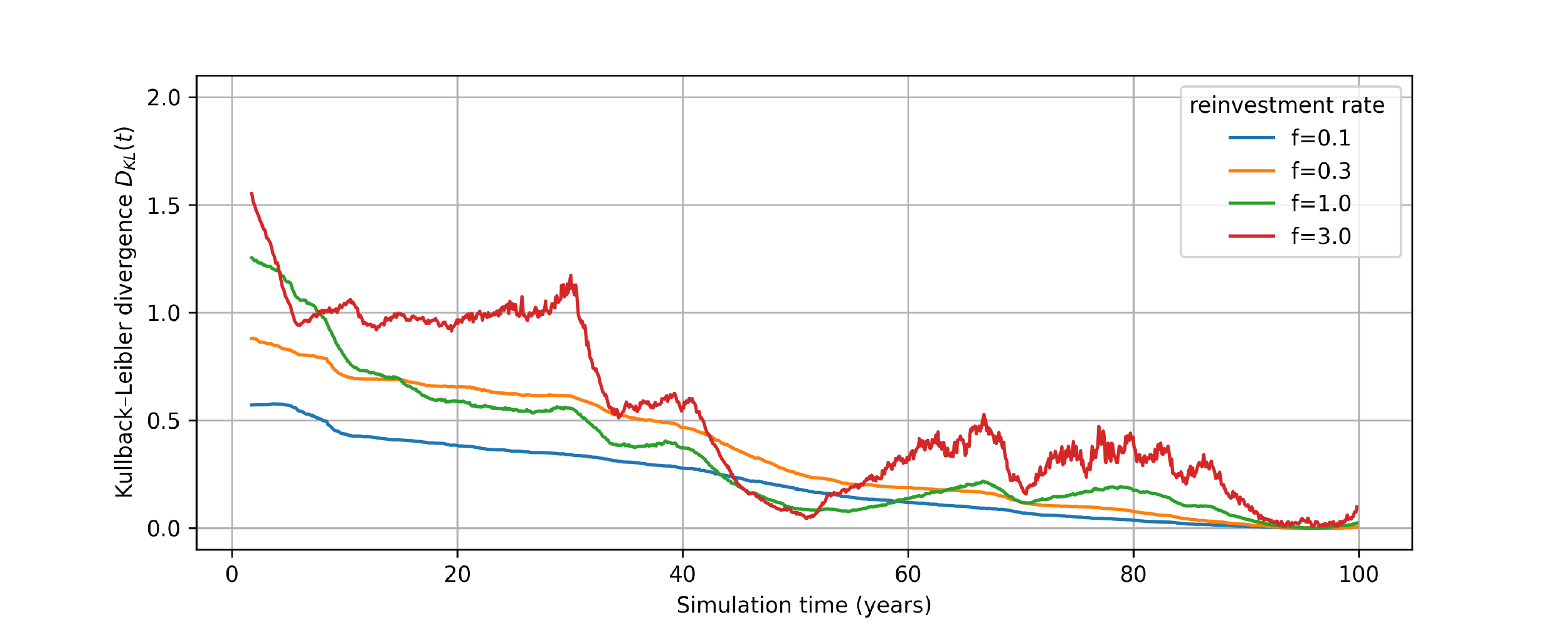}
    \caption{KL-Divergence between the Monte-Carlo distribution estimate at time $t$ versus the long-term distribution sampled over the period $t=100Y \ldots 200Y$, tested for different choices of the reinvestment rate $f$ (see: \ref{section:reinvestment}). For each parameter choice, we plot the divergence between the distribution at time $t$ to their respective long-term distribution. The choice of the parameter causes the steady-state distribution to change, and therefore the KL-divergence of the distribution of endowments at $t=0$ to this steady state has different values.}
    \label{fig:reinvestment_rate_variants}
\end{figure}

We learn that a higher rate of reinvestment leads to a mildly faster rate of convergence, however, it also leads to dynamics with a higher variance. Thus, large reinvestment rates also lead to destabilize the system. When the reinvestment rate is set to large values with $f >> 3$, it becomes increasingly likely that the market price changes on a single day are enough to bankrupt one of the funds. For example, 20\% returns on a single day for a single stock are not uncommon on the time scale of a century and a reinvestment rate of five would be enough to bankrupt a fund that has the wrong position in that single price move.

\subsection*{Results: Role of noise trader noise level}\label{section:ntv}

We repeat the same experiment as before, but now we fix $f=1.0$ and instead vary the noise trader volatility $\gamma = 0.05, 0.20, 0.40$. The result is visualized in figure~\ref{fig:divergence_nvt}, where we see that higher noise trader volatility leads to slower convergence to the steady state, and slightly larger variance during the steady state.

\begin{figure}[htb]
    \centering
    \includegraphics[width=\linewidth]{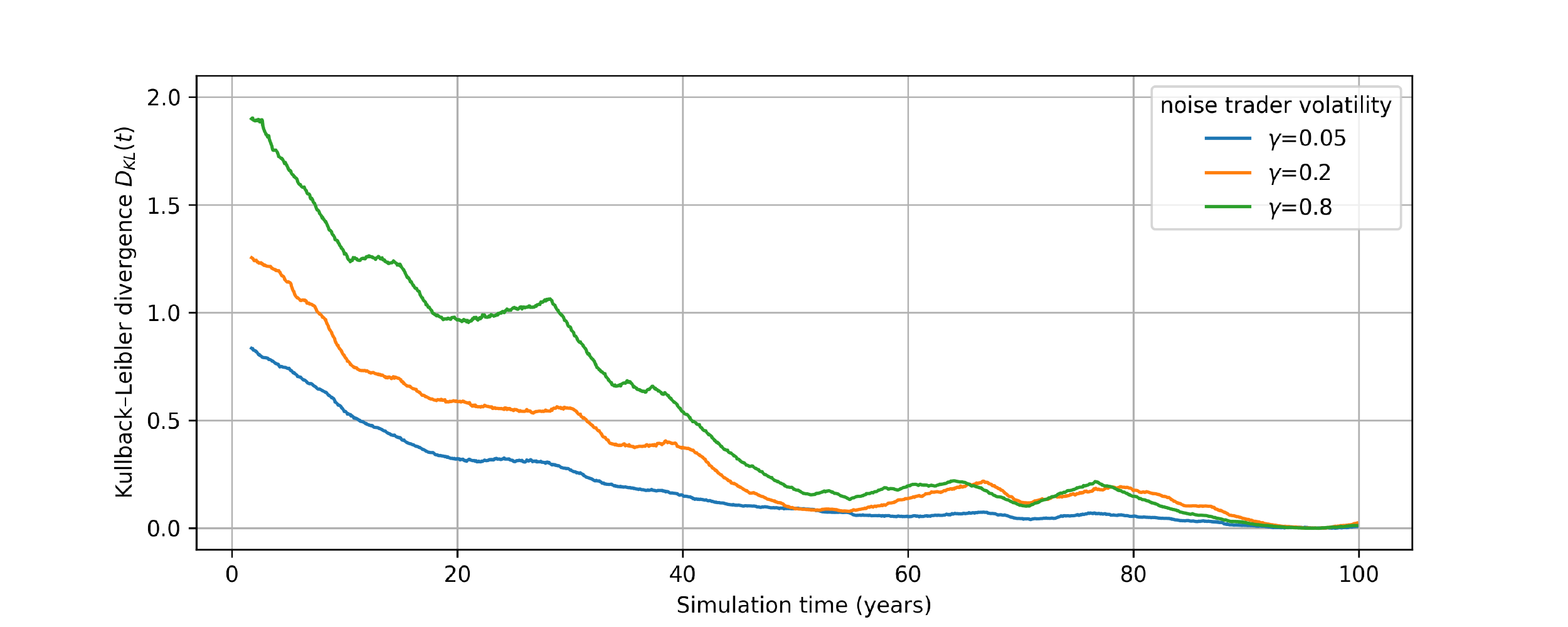}
    \caption{KL-Divergence between the state of the system at time $t$ versus the long-term distribution $t=100Y \ldots 200Y$, tested for different choices of the noise trader volatility $\gamma$ (see: \ref{section:ntv}).}
    \label{fig:divergence_nvt}
\end{figure}

\subsection*{Results: Role of fundamental value volatility}\label{section:dpv}

In this experiment, we alter the volatility of the dividend process by changing the parameter $\sigma$ from $\sigma=0.025, 0.10, 0.40$ in equation \ref{label:multivariate_geometric_brownian_motion_autocorrelation}. There appears to be no significant effect on convergence rates, and the steady state is also not altered, as is evident from figure~\ref{fig:divergence_dpv} where the trajectories of all Monte Carlo experiments are close.

\begin{figure}[htb]
    \centering
    \includegraphics[width=\linewidth]{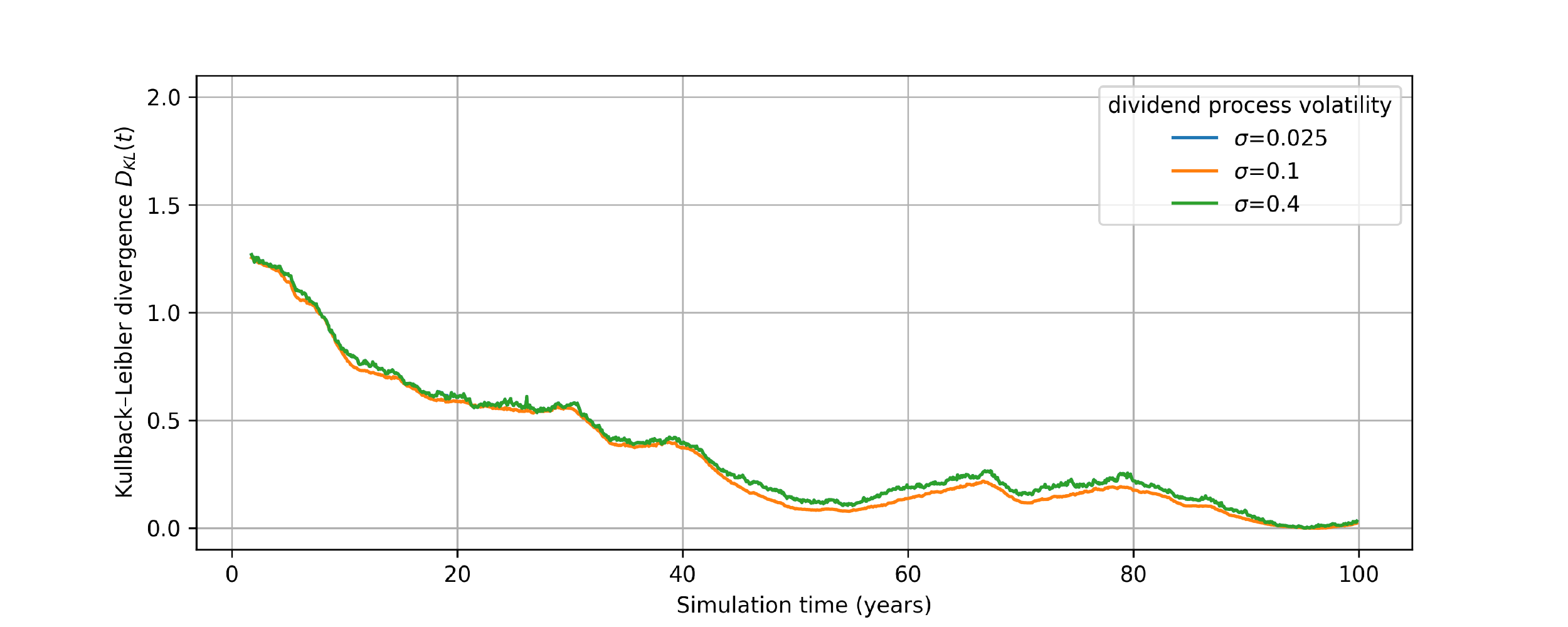}
    \caption{KL-Divergence between the state of the system at time $t$ versus the long-term distribution $t=100Y \ldots 200Y$, tested for different choices of the dividend volatility $\sigma$6 (see: \ref{section:dpv}).}
    \label{fig:divergence_dpv}
\end{figure}

\section{Kelly Criterion}

The Kelly criterion is used to determine the optimal fraction $x^*$ of wealth to invest on a risky bet. Suppose we have chosen a fraction $x$, then the wealth of our portfolio is expected to grow over one time step ahead from time $t$ from the fraction $x$ invested in the risky stock and the remainder $(x-1)$ invested in the risk-free money market instrument earning the interest rate $r$. It follows that the wealth under investment of fraction $x$ of total wealth $W(t)$ into the stock is given by 

$$
W(t+1) = x W(t) \left(\frac{p(t+1)+D(t+1)}{p(t)}-1\right) + (1-x)W(t)r
$$

\subsection*{The Investment}

Recall the fundamental value equation.
$$
    V(0) = \sum_{t}^{\infty} \frac{D(t)}{(1+k)^{t}}
$$

Suppose that the autocorrelation in dividend values is small enough to be ignored, and that the market is near efficiency. Then, by the Gordon growth model 

$$
    p(0) \approx V(0) = \frac{D(1)}{k-g},
$$

from which we learn that the relation between dividend payment and market price under those assumptions is linear. This implies if dividend's returns are normally distributed, we have normal distributed log returns $\log p(t+1)/p(t)$ on the stock. Thus, near the efficient point,  $\mathbb{E}\left[\frac{D(t+1)}{p(t)}\right] \propto k-g.$

From this approximation we learn that the returns on the stock and dividend should be highly correlated, under the assumption of little autocorrelation and fundamental-value efficient markets.

\subsection*{Dividend Process}

For convenience, we repeat here the expectation of the dividend process.
\begin{equation}\label{eqn:dividend_repeated}
\begin{split}
    \mathbb{E}\left[\frac{D(t+1)}{D(t)}\right] &= 
\exp\left(g-\frac{1}{2}\sigma^2 + \sigma\mathbb{E}[ U(t+1)]\right)\\&= 
\exp\left(g-\frac{1}{2}\sigma^2 + \sigma\omega U(t-1)\right)\\
\end{split}
\end{equation}

Recall that the expectation of the future value of the stochastic process $\sigma\mathbb{E}[ U(t+1)]$ has expectation $\sigma\omega U(t-1)$, where $U(t-1)$ can be inferred by the investor from $D(t-1)$ having correctly estimated $g$, $\sigma$ and $\omega$.

\subsection*{Solution Method}

A fully rational Kelly bettor would model possible other participants in the market and their likely influence on the returns of the investment. This quickly leads to a complicated and intractable model. We will instead introduce an approximation, based on the insight that price returns correlate highly with the dividend returns as the preceding sections showed.

Recall we use $\pi$ to denote returns to the fund, and $x$ is the fraction invested in the stock. We assume that a solution can be found of the form

\begin{equation}\label{eqn:kelly_returns_model}
\begin{split}
            \pi(x) = x \mathbb{E}\Huge[&\exp(g - \frac{\sigma^2}{2}{+}\sigma\omega U(t-1))\\
            &\cdot \exp(\mu_Y{-}\frac{\sigma_Y^2}{2}+\sigma_Y O(t))\Huge]\\ &{+}(1-x)r. \\
    \end{split}
\end{equation}

In the first term in the product we recognize the dividend process from equation \ref{eqn:dividend_repeated}. The second component is due to an approximation $Y$ of the stock returns. This process $Y$ has two parameters $\mu_Y$, the drift, and $\sigma_Y$, the volatility, that are dynamically estimated by the Kelly investor. The process is defined as:

\begin{equation}
    \frac{Y(t+1)}{Y(t)} = \exp\left(\mu_Y - \frac{1}{2}\sigma_Y^2 + \sigma_Y O(t) \right)
\end{equation}

with $O(t)$ a standard normal variable. To summarize, the approximation that we make is that

$$
\log \left(\frac{p(t+1)}{p(t)} -1 + \frac{D(t+1)}{p(t)} \right) \approx \log\left(\frac{D(t+1)}{D(t)} \cdot \frac{Y(t+1)}{Y(t)}  \right).
$$

We test whether this approximation yields the optimal investment fraction in a Monte-Carlo experiment with dividends and prices from the main model. The fraction given by the approximation will be $x^*$, and we will compare this with other fractions $x$. In Figure~\ref{fig:kelly_numerical_test} we plot on the x-axis the different fractions invested $x/x^*$, and on the y-axis the expected returns. The figure shows that the highest mean returns are obtained near $x^*=x$, meaning the approximation is close to optimal.  

\subsection*{Kelly Strategy}\label{section:kelly_strat}

The Kelly bettor maximizes log returns in equation~\ref{eqn:kelly_returns_model}. This gives the following optimization problem:

\begin{equation*}
\begin{split}
    \max_x  \pi(x) &= x\left(g{+}\mu_Y{+}\sigma\omega U(t{-}1){-}r{-} 2\rho\sigma\sigma_Y\right) {-}\frac{1}{2}(x\sigma)^2 {-}\frac{1}{2}(x\sigma_Y)^2\\
    &\text{subject to}
    {-}\lambda^{*} \leq x \leq \lambda^{*},\\
\end{split}
\end{equation*}
wherein $\rho$ is the estimated correlation coefficient between the market prices and dividends. This is maximized when:

\begin{equation}\label{eqn:kelly_dividends_optimal}
x^{*} = \max\left(-\lambda^*, \min\left(\lambda^*,  \frac{g + \mu_Y + \sigma\omega U(t-1) - r - 2\rho\sigma\sigma_y}{\sigma^2 + \sigma_y^2} \right)\right).
\end{equation}

We will assume the Kelly bettor has no information other than the dividend announcements and market prices. Therefore, the Kelly bettor will have at time $t$ an estimate of the parameters of equation~\ref{eqn:kelly_dividends_optimal} based on the public information up to time $t$.

\begin{figure}
    \centering
    \includegraphics[width=0.9\linewidth]{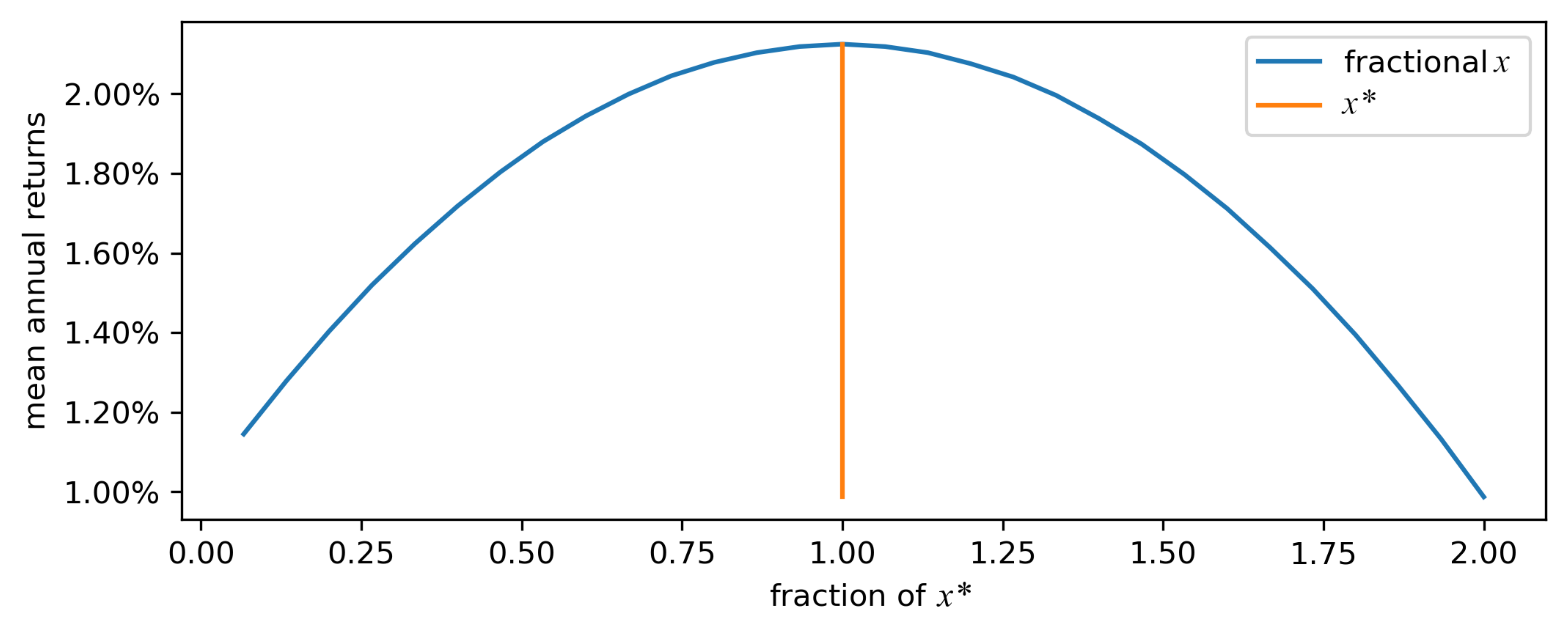}
    \caption{A Monte-Carlo experiment illustrates that the approximate Kelly strategy derived in section \ref{section:kelly_strat} is numerically close to the optimal strategy.}
    \label{fig:kelly_numerical_test}
\end{figure}

\subsection*{Survival}

In previous studies on markets with different constraints than ours \cite{Hens2009}, it is shown that the Kelly criterion strategy is a survival strategy.
In our discrete time model, we observe that the Kelly criterion is indeed highly likely to survive for long time spans, but still it is possible that the Kelly bettor fails. One scenario in which the Kelly bettor, using equation~\ref{eqn:kelly_dividends_optimal} to invest, can go extinct, is when for a long run market prices increase, followed by a sudden drop. During the build-up phase, the Kelly bettor estimates large values for $g$ and $\mu_Y$, and aggressively uses leverage. Depending on how the investor updates the parameter estimates, a sudden drop in prices and or dividends may happen fast enough that the Kelly bettor still maintains positive leverage, leading to excessive losses and bankruptcy. The different possibilities for failure of one or more strategies also indicate that the system is more unstable with the strategy added (see figure~\ref{fig:kelly_survivability}).

\begin{figure}
    \centering
    \includegraphics[width=0.9\linewidth]{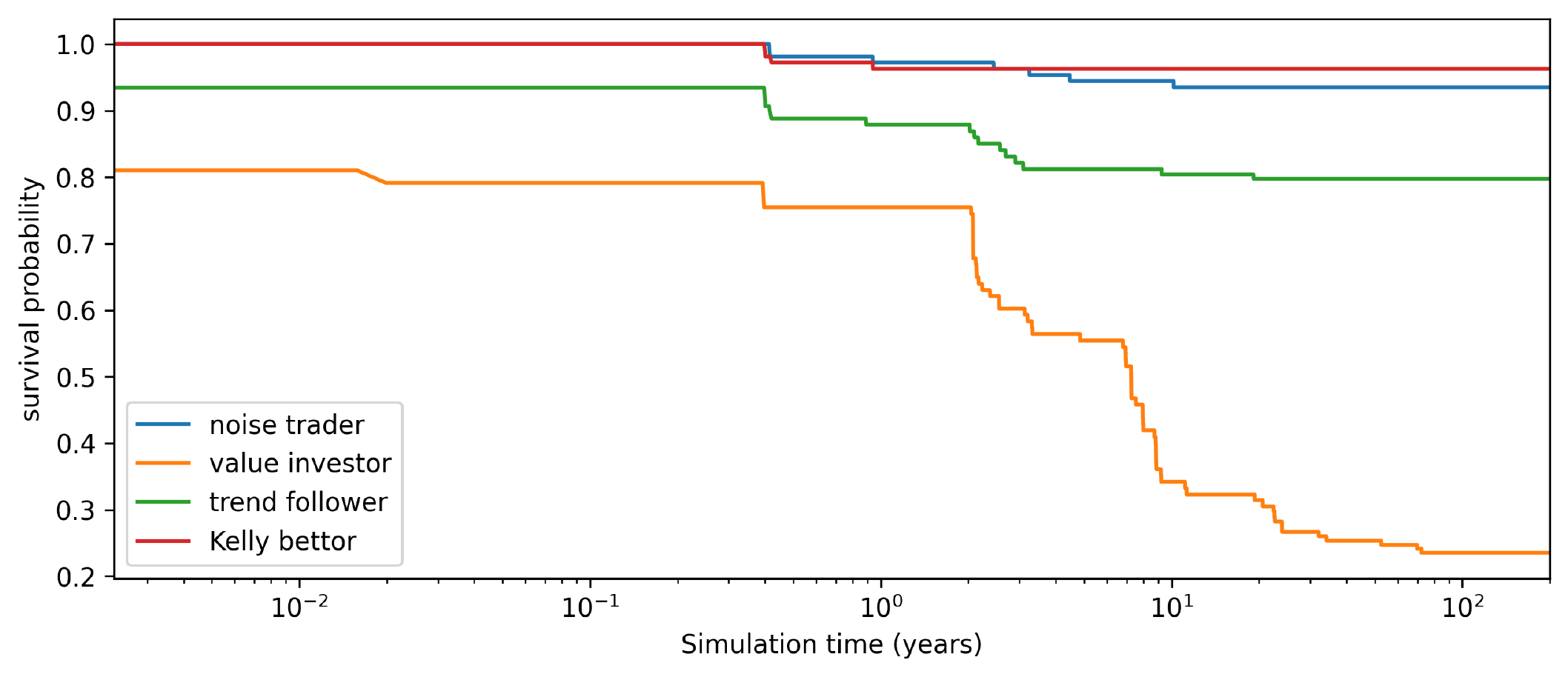}
    \caption{This plot shows the estimated probability that a fund with a given strategy still has positive wealth at the time $t$, out of a Monte-Carlo experiment with 5000 samples. }
    \label{fig:kelly_survivability}
\end{figure}

\fi

\end{document}